\documentclass[12pt,preprint]{aastex}

\begin{document}

\title{Exploring the Variable Sky with LINEAR. I. Photometric Recalibration with SDSS} 

\author{
Branimir Sesar\altaffilmark{\ref{Caltech},\ref{Washington}},
J.~Scott Stuart\altaffilmark{\ref{LLMIT}},
\v{Z}eljko Ivezi\'{c}\altaffilmark{\ref{Washington}},
Dylan P.~Morgan\altaffilmark{\ref{Washington},\ref{BU}},
Andrew C. Becker\altaffilmark{\ref{Washington}},
Przemys\l aw Wo\'zniak\altaffilmark{\ref{LANL}}
}

\altaffiltext{1}{Division of Physics, Mathematics and Astronomy, Caltech,
                 Pasadena, CA 91125\label{Caltech}}
\altaffiltext{2}{Lincoln Laboratory, Massachusetts Institute of Technology,
                 244 Wood Street, Lexington, MA 02420-9108\label{LLMIT}}
\altaffiltext{3}{Department of Astronomy, University of Washington, P.O.~Box
                 351580, Seattle, WA 98195-1580\label{Washington}}
\altaffiltext{4}{Los Alamos National Laboratory, 30 Bikini Atoll Rd.,
                 Los Alamos, NM 87545-0001\label{LANL}}
\altaffiltext{5}{Department of Astronomy, Boston University, 725 Commonwealth
                 Avenue, Boston, MA 02139\label{BU}}

\begin{abstract}
We describe photometric recalibration of data obtained by the asteroid survey
LINEAR. Although LINEAR was designed for astrometric discovery of moving
objects, the dataset described here contains over 5 billion photometric
measurements for about 25 million objects, mostly stars. We use SDSS data from
the overlapping $\sim$10,000 deg$^2$ of sky to recalibrate LINEAR photometry,
and achieve errors of 0.03 mag for sources not limited by photon statistics,
with errors of 0.2 mag at $r\sim18$. With its 200 observations per object on
average, LINEAR data provide time domain information for the brightest 4
magnitudes of SDSS survey. At the same time, LINEAR extends the deepest similar
wide-area variability survey, the Northern Sky Variability Survey, by 3 mag. We
briefly discuss the properties of about 7,000 visually confirmed periodic
variables, dominated by roughly equal fractions of RR Lyrae stars and eclipsing 
binary stars, and analyze their distribution in optical and infra-red
color-color diagrams. The LINEAR dataset is publicly available from the SkyDOT
website (\url{http://skydot.lanl.gov}).
\end{abstract}


\section{Introduction\label{introduction}}

Variability is an important phenomenon in astrophysical studies of structure and
evolution, both stellar, galactic, and extragalactic. Some variable stars, such
as RR Lyrae stars, are an excellent tool for studying the Galaxy. Being nearly
standard candles (thus making distance determination relatively straightforward)
and being intrinsically bright, they are a particularly suitable tracer of
Galactic structure (e.g., \citealt{ses07}, and references therein). Similarly,
eclipsing binaries can be used as distance indicators \citep{gui98}, are
excellent probes of stellar physics \citep{pz05}, and offer a unique method for
measuring stellar masses and other parameters as well (e.g.,
\citealt{and91,tag10}).

Despite the importance of variability, the variable optical sky remains largely
unexplored and poorly quantified, especially at the faint end ($V>15$). To what
degree different variable populations contribute to the overall variability, how
they are distributed in magnitude and color, and what are the characteristic
timescales and dominant mechanisms of variability, are just some of the
questions that still remain to be answered. To address these questions, several
contemporary projects aimed at regular monitoring of the optical sky were
started. The four most prominent wide-area (more than a few thousand deg$^2$)
variability surveys in terms of depth and cadence are as follows:
\begin{itemize}
\item
ROTSE-I \citep{ake00} monitored the entire observable sky twice a night from
$V=10$ to a limit of $V=15.5$. The Northern Sky Variability Survey (NSVS;
\citealt{woz04}) is based on ROTSE-I data and contains light curves for about 14
million objects, with 100-500 measurements per object collected over one year.
Typical photometric precision of this dataset is about 0.02 mag for sources not
limited by photon statistics.
\item
The All Sky Automated Survey (ASAS; \citealt{poj02}) monitors the entire
southern and part of the northern sky ($\delta<+28\arcdeg$) to a limit of
$V=14$, including about 15 million stars. Typical photometric precision is about
0.05 mag. The third release of the ASAS Catalog of Variable Stars contains close
to 50,000 variable stars. About 80\% of these are new discoveries (not
previously cataloged).
\item
The Lowell Observatory Near Earth Objects Survey Phase I (LONEOS-I;
\citealt{mic08}) provides photometric data for 1430 deg$^2$ of northern sky that
has been imaged at least 28 times between 1998 and 2000. The LONEOS-I camera
used no bandpass filter and reached a depth of $R\sim18.5$. Typical photometric
precision of this dataset is about 0.02 mag for sources not limited by photon
statistics.
\item
The Palomar Transient Factory (PTF;\citealt{law09} and \citealt{rau09}) is an
ongoing wide-area, two-band (SDSS-$g^\prime$ and Mould-$R$ filters), deep
($R\sim20.6$, $g^\prime\sim21.3$) survey aimed at systematic exploration of the
optical transient sky. As of Fall 2011, PTF has observed about 7300 deg$^2$ of
northern sky at least 30 times in the Mould-$R$ band ($\sim1800$~deg$^2$ of sky
with more than 100 epochs). Typical photometric precision of this dataset is
better than 0.01 for sources not limited by photon statistics.
\end{itemize}
A comprehensive review of past and ongoing variability surveys can be found in
\citet{bec04}.

With hundreds of observations per object, the NSVS and ASAS datasets represent
unique resources. Although a number of massive synoptic surveys, such as
Pan-STARRS \citep{kai02} and Large Synoptic Survey Telescope (LSST;
\citealt{ive08}), will significantly improve our knowledge of the variable faint
sky, they will not obtain hundreds of observations per object any time soon.

A downside of NSVS, ASAS, LONEOS, and PTF datasets is the lack of precise
multi-color photometry, which can provide valuable information about sources in
addition to light curve characteristics \citep{cov07}. The Sloan Digital Sky
Survey (SDSS; \citealt{yor00}) has obtained five-band photometry for hundreds of
millions of stars, but its temporal coverage is very poor. The only region on
the sky with more than ten observations per source is the $\sim$300 deg$^2$
large equatorial region called Stripe 82 (\citealt{ses07}, and references
therein). At the same time, NSVS and ASAS are not deep enough to make efficient
use of SDSS photometry. This goal to combine a large-area variability survey
providing hundreds of observations with exquisite SDSS photometry was recently
made reachable by the LINEAR survey.

MIT Lincoln Laboratory (MITLL) has operated the Lincoln Near-Earth Asteroid
Research (LINEAR) program since 1998 to search for asteroids. The focus of the
program is to discover and track near-Earth asteroids (NEAs) larger than 1 km
diameter, but LINEAR has also discovered and tracked many main-belt asteroids
and smaller NEAs. Since late 2002, the LINEAR program has maintained an archive
of all of its imagery data. This archive now contains approximately 6 million
images of the sky, most of which are 5 megapixel images covering 2 deg$^2$. The
LINEAR image archive contains a unique combination of depth ($V<18$), coverage
area, and cadence that is potentially useful in exploring a number of
time-varying astronomical phenomena. 

Since LINEAR project was focused on astrometric survey, standard stars required
for precise photometric calibration were often unavailable. Furthermore, a
significant fraction of data was taken in non-photometric conditions. The main
purpose of this paper is to describe photometric recalibration of LINEAR data
using stars observed by SDSS as photometric standards. Companion papers are
discussed in \S~\ref{summary}.

In \S~\ref{survey} we describe LINEAR survey in more detail, and discuss
recalibration procedures in \S~\ref{recalib}. The recalibrated dataset is
characterized in \S~\ref{characterization} and a preliminary analysis of about
7,000 confirmed periodic variables is presented in \S~\ref{analysis}. Our
results are summarized in \S~\ref{summary}. 

\section{LINEAR Survey\label{survey}}

\subsection{Hardware Characteristics}

The LINEAR program operates two telescopes at the Experimental Test Site located
within the US Army White Sands Missile Range in central New Mexico, at an
altitude of 1,506 m above sea level, and a latitude of $33.8\arcdeg N$. The
program uses two, essentially identical, equatorial-mounted telescopes of the
GEODSS (Ground-based Electro-Optical Deep Space Surveillance) type, usually
denoted L1 and L2. Each telescope is a folded prime-focus design with a primary mirror diameter of 1.016 m diameter, and $f/\#$ of 2.15 \citep{sto00}.

\begin{figure}
\epsscale{1.0}
\plotone{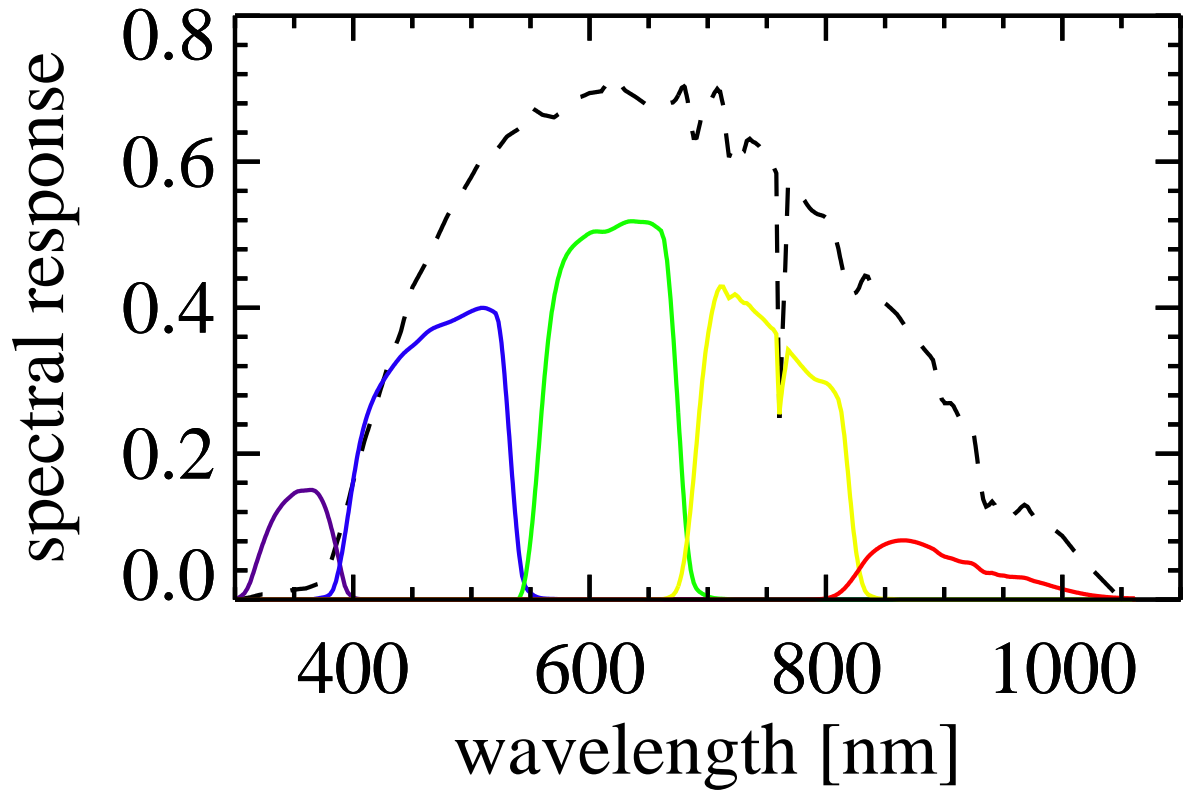}
\caption{
The LINEAR ({\em dashed line}) and SDSS $ugriz$ spectral response curves
({\em solid lines}). The response curve for LINEAR includes the spectral quantum
efficiency of the back-illuminated CCD, reflectivity of the aluminum mirrors,
and atmospheric transmissivity at airmass of 1.3. The response curves for SDSS
include the same atmospheric absorption.
\label{response_curves}}
\end{figure}

The two telescopes of the LINEAR project are each equipped with a CCD camera
developed at MITLL using CCDs developed and manufactured by MITLL for the US Air
Force GEODSS system \citep{bur98}. These CCID-16 sensors are 5 megapixel
(2560x1960), back-illuminated, frame-transfer CCDs with fast readout, low dark
current, and low read noise. The CCID-16 sensors have 8 readout regions
(hereafter referred to as cells) and have high quantum efficiency over a broad
part of the visible and near-infrared spectrum (with overall solar-weighted
quantum efficiency of 65\% and peak efficiency of 96\% at 620 nm, see
Figure~\ref{response_curves}). The cameras have no spectral filters. The
combination of the CCID-16 camera and the 1-m GEODSS telescope produces an
instantaneous field-of-view of $1.60\arcdeg$ by $1.23\arcdeg$ ($\sim2$ deg$^2$),
with $2\arcsec.25$ pixels. In Section~\ref{recalib}, the combination of CCID-16
sensors and L1 and L2 telescopes will be referred to as the L1 and L2 setups.

For a few months in 2003 (approximately August to November) the camera on the L2
telescope was replaced with a very similar camera containing a smaller format
sensor called the CCID-10. This smaller sensor is identical to the CCID-16s in
terms of manufacture and processing, readout electronics and settings, spectral
quantum efficiency and pixel size. The only difference between the smaller
format camera and the larger format camera is the number of pixels, 1024x1024
for the small camera, and the number of readout regions (4 cells). During the
four months in 2003 that L2 was operating with the CCID-10 camera, the
field-of-view was about $0.64\arcdeg$ by $0.64\arcdeg$ ($\sim0.4$ deg$^2$). In
Section~\ref{recalib}, the combination of CCID-10 sensor and L2 telescope will
be referred to as the small setup.

\subsection{Observing Conditions and Cadence}

The telescope site has reasonably dark skies, approaching $\mu_V=21$ mag
arcsec$^{-2}$ on the best nights. However, the LINEAR program operates on nights
with suboptimal conditions including haze, bright moonlight (to within 3 days of
the full moon), scattered clouds, and airglow. LINEAR images are frequently
taken under skies as bright as $\mu_V=16$ mag arcsec$^{-2}$. The point spread
function includes unknown contributions from atmospheric seeing and various
system contributions (e.g., defocus and dome airflow) and is typically around
$5\arcsec$ FWHM, resulting in slightly undersampled images.

The image integration time is allowed to vary over the seasons so that a roughly
constant number of search fields fills the available dark time. During the
earlier days of the LINEAR program (roughly 1998 to 2006), the integration time
was 10-12 seconds on long winter nights and 3-5 seconds on short summer nights.
In recent years (roughly 2007 to 2009), the integration time was increased to
15-18 seconds in the winter and 8-13 seconds in the summer. Under these
observing conditions, the magnitude for which the magnitude error is $\sim0.2$
mag (the $5-\sigma$ detection limit) is about 18 mag on average.

\begin{figure}
\epsscale{1.0}
\plotone{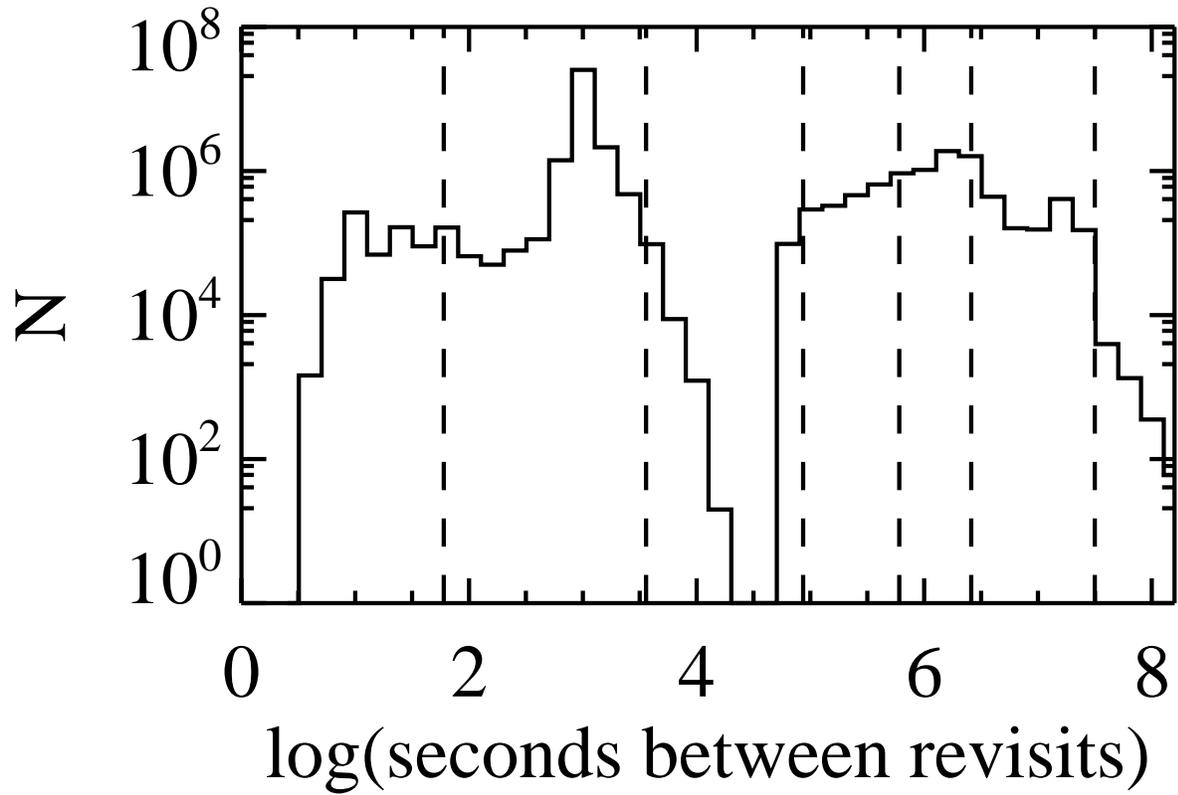}
\caption{
The cadence of LINEAR observations measured as the time between revisits of the
same patch of sky. From left to right, the dashed lines show 1-minute, 1-hour,
1-day, 1-week, 1-month, and 1-year cadences. The LINEAR temporal coverage is
fairly uniform, with a peak at $\sim1000$ seconds corresponding to the main
15-min cadence.
\label{cadence}}
\end{figure}

As shown in Figure~\ref{cadence}, the cadence of LINEAR observations (time
between revisits of the same patch of sky) is fairly uniform, with a peak at
$\sim15$ minutes and a gap at $\sim8$ hours. A more detailed description of
LINEAR observing strategy is given in Appendix~\ref{appendix:cadence}.

\subsection{Photometry and Astrometry using SExtractor}

The LINEAR project has been archiving all of its image data since December 2002
through the present. The data used for this work encompass all imagery collected
from December 2002 through March 2008. After selecting the LINEAR imagery data
that overlaps the main SDSS survey region (i.e.~galactic latitudes greater than
$30\arcdeg$) and SDSS stripe 82 region
($20^h 32^m <\alpha_{J2000.0} < 04^h 00^m$,
$-1.26\arcdeg<\delta_{J2000.0} < +1.26\arcdeg$, $\sim280$ deg$^2$)), the data
comprises about 1.8 million images. The bias and flat-field corrections were
applied to these images following standard methodology. 

The extraction of sources from LINEAR images and the creation of source catalogs
were done using the SExtractor software (\citealt{ba96}; see the SExtractor
users manual\footnote{Available at \url{https://www.astromatic.net/pubsvn/software/sextractor/trunk/doc/sextractor.pdf}} for more details on the source
detection and measurement process). The preliminary astrometry for extracted
sources was obtained by converting centroid pixel coordinates into equatorial
J2000.0 right ascension (R.A.) and declination (Dec) by application of the
gnomonic projection and a-priori telescope pointing information. This projection
does not correct for telescope pointing errors on the order of several pixels,
telescope optical distortion on the order of a few pixels near the corners of
the image, or differential refraction on the order of a few pixels at typical
airmasses. When compared to the USNO-B catalog \citep{mon03} astrometry, the
average uncertainty in the preliminary astrometry is about $\sim6\arcsec$ in
R.A.~and Dec coordinates. Therefore, the astrometry reported by SExtractor must
be recalibrated before matching extracted LINEAR sources against other catalogs.
The recalibration of preliminary astrometry is described in the next section.

The preliminary fixed-aperture photometry for each source is computed by summing
the pixels within a 5 pixel ($\sim11\arcsec$) radius circular aperture and
subtracting the background estimate for each pixel computed in the background
estimate step. The uncertainty on the computed flux is computed assuming Poisson
statistics in detected electrons with a gain conversion factor of $2.5 e^-/ADU$.
The pixels with counts above 17000 were not used as the CCDs begin to experience
non-linear response above this level (corresponding to 42000 electrons), and
they saturate at full-well capacity of 75,000 electrons.

\section{Recalibration of LINEAR astrometry and photometry\label{recalib}}

In this Section, we describe the astrometric and photometric recalibration of
LINEAR data and quantify the quality of recalibrated data. The recalibration and
data quality assessment are done on a field-by-field basis, where a field is
defined as part of a LINEAR image observed by a single CCD cell (there are 4
cells in the small setup and 8 cells in the L1 and L2 setups). The only exception to this
is the astrometric recalibration, which is done on a image-by-image basis.
However, the quality of astrometric recalibration is still assessed on a
field-by-field basis.

\subsection{Astrometric recalibration\label{astro_recal}}

We recalibrate the preliminary astrometry using the Astrometry.net software
\citep{lan10}. The Astrometry.net uses the $x$ and $y$ (hereafter, pixel)
positions of sources extracted by SExtractor and looks for ``skymarks'' (sets of
four stars) that correspond to pre-computed ``skymarks'' found in the ``clean''
USNO-B catalog \citep{bar08}. Once a skymark is found the code obtains an
initial astrometric solution (a transformation from the pixel coordinates into
the J2000.0 Equatorial system) in the form of a third-degree polynomial in $x$
and $y$ coordinates. Using this initial astrometric solution the code looks for
additional reference USNO-B stars in the image, and when it finds them it
further refines the initial astrometric solution.

\begin{figure}
\epsscale{0.7}
\plotone{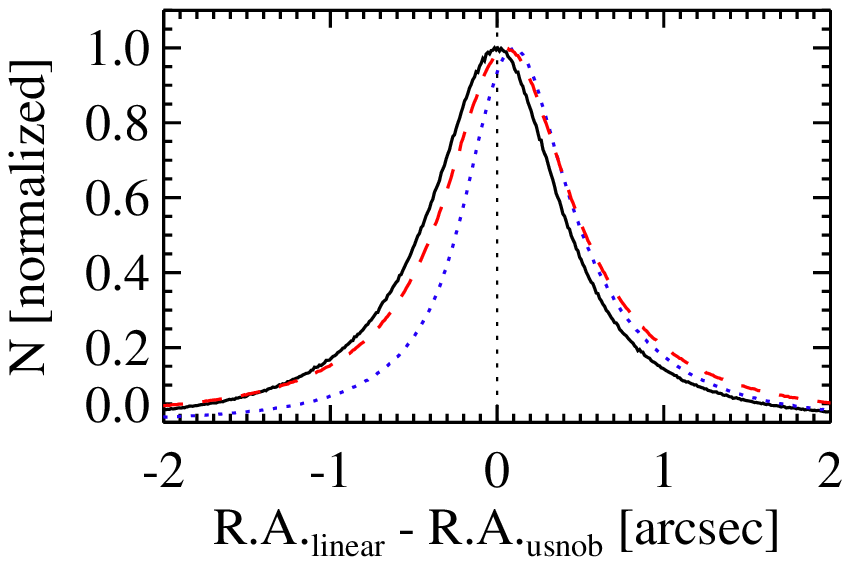}

\plotone{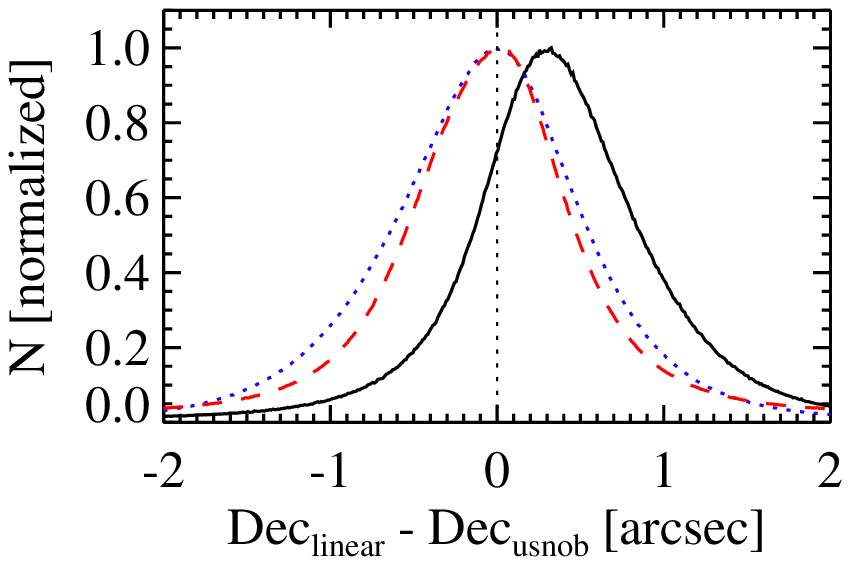}
\caption{
A comparison of USNO-B and recalibrated LINEAR R.A.~({\em top}) and Dec
single-epoch coordinates ({\em bottom}) for the small ({\em solid line}), L1
({\em dotted line}), and L2 setup ({\em dashed line}). The width of the
distributions (determined from the interquartile range) puts the astrometric
precision in LINEAR single-epoch coordinates at $\sim0.6\arcsec$. The systematic
offsets in LINEAR coordinates are $\la0.3\arcsec$ and have been subsequently
corrected for. The dotted line at $0\arcsec$ was added to guide the eye.
\label{astrometry_comparison}}
\end{figure}

The quality of recalibrated LINEAR astrometry is estimated by comparing the
LINEAR positions with reference USNO-B positions
(Figure~\ref{astrometry_comparison}). We find that the recalibrated LINEAR
single-epoch coordinates have an rms scatter of $\sim0.6\arcsec$ and are offset
$\la0.3\arcsec$ with respect to the USNO-B coordinates. These systematic offsets
are independent of pixel positions or the position of the image on the sky, and
are subtracted from recalibrated LINEAR coordinates. Using a subset of data, we
have found that the offsets decrease if the Astrometry.net is allowed to search
for even more reference USNO-B stars when refining the initial astrometric
solution. 

To find fields with bad astrometry, we calculate the median and rms scatter of
astrometric residuals (difference between LINEAR and USNO-B positions) in
R.A.~and Dec for each field, and tag fields as bad if the median or rms scatter
in either coordinate (R.A.~or Dec) is greater than $2\arcsec$. We find that
about 7\% of all fields have bad astrometry, with the majority of bad fields
($\sim90\%$) originating from the L2 setup, and remove them from further
processing. Visual inspection of images containing bad fields revealed that
bad fields lack several pixel columns, which were most probably dropped due to
cell readout problems.

In order to recalibrate the LINEAR photometry, we positionally matched the
LINEAR dataset to the Sloan Digital Sky Survey Data Release 7 (SDSS DR7;
\citealt{aba09}) imaging catalog. The SDSS DR7 imaging catalog provides
homogeneous and deep ($r<22.5$) 1-2\% accurate photometry in five band-passes
($u$, $g$, $r$, $i$, and $z$) of more than 11,000 deg$^2$ of the North Galactic
Cap, and three stripes in the South Galactic Cap totaling 740 deg$^2$. The SDSS
positions are accurate to better than $0.1\arcsec$ per coordinate (rms) for
sources with $r<20.5$ \citep{pie03}, and the morphological information from the
images allows reliable star-galaxy separation to $r\sim21.5$ \citep{lup02}.

We match LINEAR sources to SDSS DR7 non-saturated, ``primary''
objects\footnote{See \url{http://cas.sdss.org/dr6/en/help/docs/algorithm.asp?key=flags}}
using a $3\arcsec$ radius, and find $\sim4.8$ billion matched pairs (there are
24 million unique SDSS sources included in these pairs). There are $\sim1.6$
billion LINEAR sources without a SDSS match (``orphans''), and they include
observations associated with stars saturated in SDSS ($r<13$, $\sim60\%$ of
orphans), image artifacts (e.g., due to charge bleeding and diffraction spikes),
moving objects such as asteroids, and other transients.

\subsection{Photometric recalibration}

We model the recalibrated LINEAR magnitudes as
\begin{equation}
m_{linear} = \alpha m_{instr} + C(g-i) + Z(x,y),\label{mlinear}
\end{equation}
where $\alpha$ measures non-linearity of SExtractor-supplied instrumental
magnitudes $m_{instr}=-2.5\log(counts)$, $C$ is a color-dependent term that
accounts for the per-exposure change in the effective central wavelength of
the LINEAR bandpass, and $Z(x,y)$ is our model for the field-specific magnitude
zero-point dependent on the $x-y$ position in the image, defined as
\begin{equation}
Z(x,y) = Z_0 + Z_1x + Z_2y + F(x,y)\label{zero_point},
\end{equation}
where $Z_0$, $Z_1$, and $Z_2$ vary from field to field, and $F(x,y)$ is the
``super flat-field'' (fixed for each observational setup). Note that the
color-independent airmass term (gray extinction) is implicitly included in
$Z_0$.

To recalibrate the LINEAR photometry we need a catalog of photometric standard
stars in the LINEAR photometric (unfiltered) system. Since such a catalog does
not exist, we use the SDSS imaging catalog and utilize SDSS $gri$ photometry to
derive magnitudes in the LINEAR photometric system.

\subsubsection{LINEAR Magnitudes Synthesized from SDSS Photometry}

We define LINEAR magnitudes synthesized from the SDSS photometry as
\begin{equation}
m_{sdss} = r + f(g-i), \label{msdss1}
\end{equation}
where $f(g-i)$ is some function of $g-i$, and SDSS $gri$ magnitudes are
{\em not} corrected for the ISM extinction.

We find $f(g-i)$ iteratively by first adopting $\alpha=1.0$ in
Equation~\ref{mlinear}, $F(x,y) = 0$ in Equation~\ref{zero_point}, and
$f(g-i)=0$ in Equation~\ref{msdss1}. For each field, we determine $Z_0$, $Z_1$
and $Z_2$ by minimizing
\begin{equation}
\chi^2 = \Sigma [(m_{linear}-m_{sdss})/\sigma_{linear}]^2\label{chi_min},
\end{equation}
where $\sigma_{linear}$ is the magnitude error supplied by the SExtractor
(errors in $m_{sdss}$ are assumed to be much smaller than errors in
$m_{linear}$).

\begin{figure}
\epsscale{1.0}
\plotone{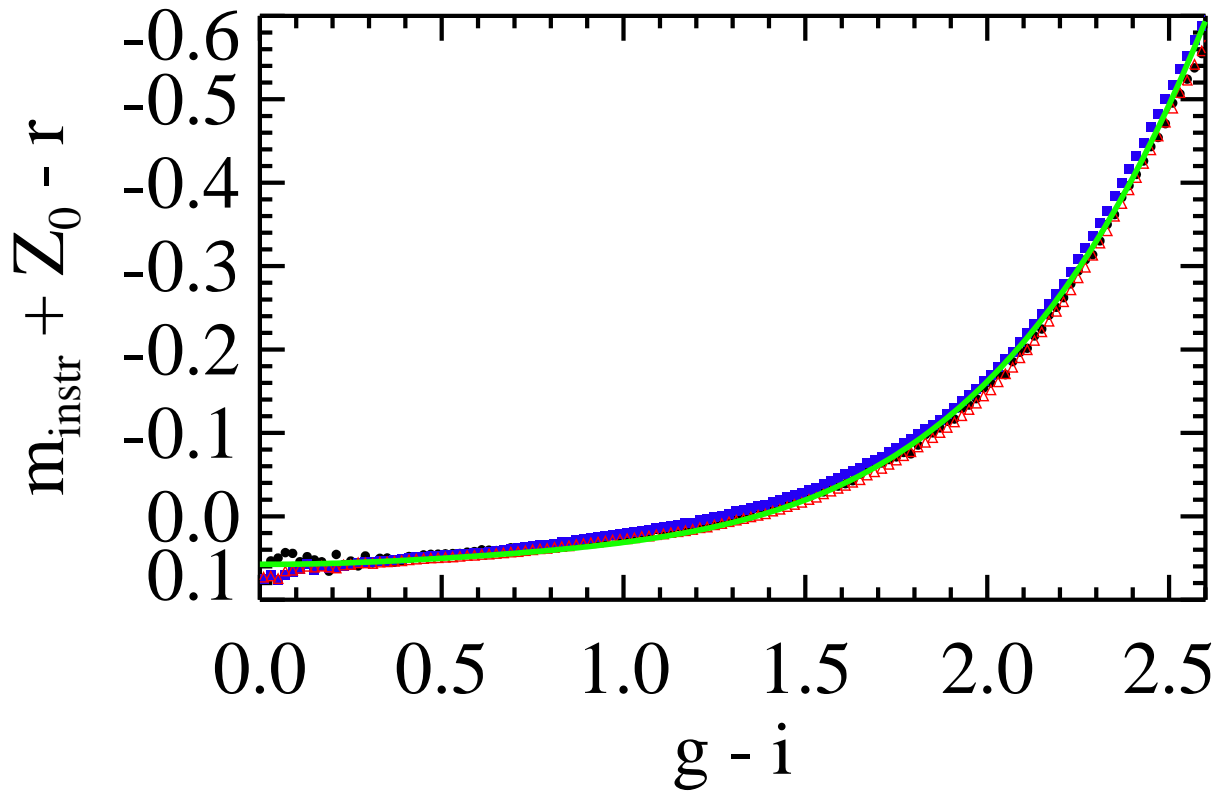}
\caption{
The color term between the zero-point corrected LINEAR magnitude,
$m_{instr}+Z_0$, and SDSS $r$-band magnitude for the small ({\em dots}), L1
({\em solid squares}), and L2 setup ({\em open triangles}). The solid line shows
the best fit for all three setups. The rms scatter between the best fit for all
three setups and fits to individual setups is smaller than 0.01 mag. The SDSS
$g-i$ color is not corrected for ISM extinction.
\label{m_sdss}}
\end{figure}

The LINEAR observations of calibration stars must be matched within $2\arcsec$
to isolated\footnote{SDSS CHILD flag set to 0}, unresolved, SDSS objects
brighter than $r=17$, with $0 < g-i < 2.6$. Having $Z(x,y)$ for each field, we
calculate $m_{instr}+Z(x,y)-r$ residuals for all calibration stars, bin the
values in $g-i$ bins, and plot the median residuals in Figure~\ref{m_sdss} for
the small, L1, and L2 setup. We plot medians for each setup separately because
each setup represents a separate optical, and therefore photometric, system.

Even though the three setups are different, their residuals depend similarly on
$g-i$ (rms scatter around the mean is 0.01 mag), as indicated by medians plotted
in Figure~\ref{m_sdss}. We therefore fit a fourth-degree polynomial to all
medians, and obtain
\begin{equation}
f(g-i) = 0.0574 + 0.004(g-i) - 0.056 (g-i)^2 + 0.052 (g-i)^3 -0.0262(g-i)^4.
\end{equation}
The transformation from SDSS $gri$ to LINEAR magnitudes, as defined by
Equation~\ref{msdss1}, is then
\begin{equation}
m_{sdss} = r + 0.0574 + 0.004(g-i) - 0.056 (g-i)^2 + 0.052 (g-i)^3 -0.0262(g-i)^4\label{msdss}.
\end{equation}
This equation effectively turns the SDSS imaging catalog into a catalog of
LINEAR photometric standard stars.

\subsubsection{Super Flat-fields and Non-linearity of Instrumental Magnitudes}

\begin{figure}
\epsscale{0.84}
\plotone{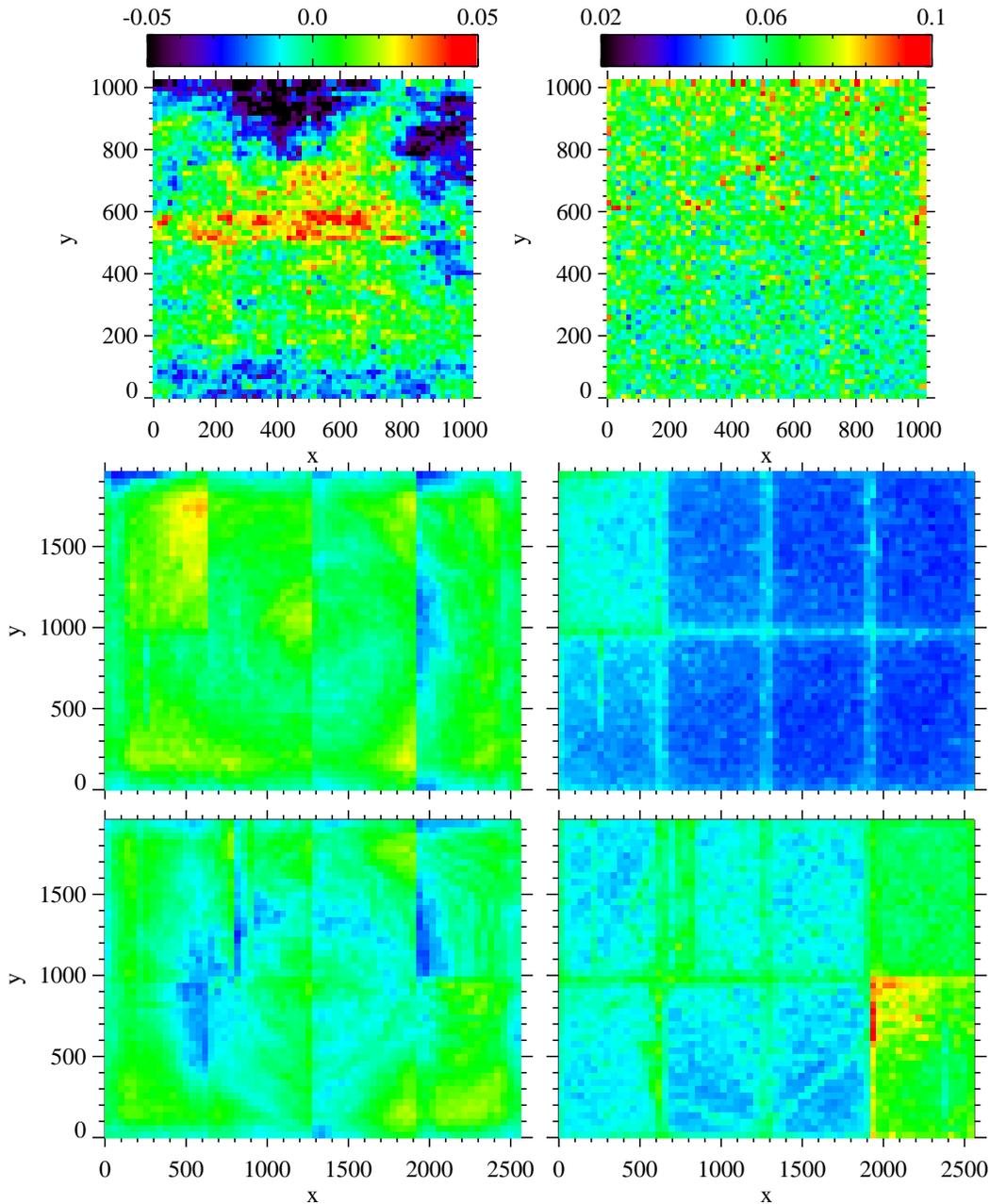}
\caption{
The distribution of $m_{linear}-m_{sdss}$ residuals binned in the $x-y$ plane
for the small ({\em top}), L1 ({\em middle}), and L2 setup ({\em bottom}). The
panels on the left show median values in each bin, and the panels on the right
show rms scatter in a bin. The values are color-coded according to legends, with
values outside the range saturating. The patterns seen on the left are magnitude
zero-point variations that were not taken out when LINEAR images were
flat-fielded. Therefore, the maps on the left represent ``super flat-fields''
that need to be subtracted during recalibration.
\label{xy_map_nosp}}
\end{figure}

\begin{figure}
\epsscale{0.85}
\plotone{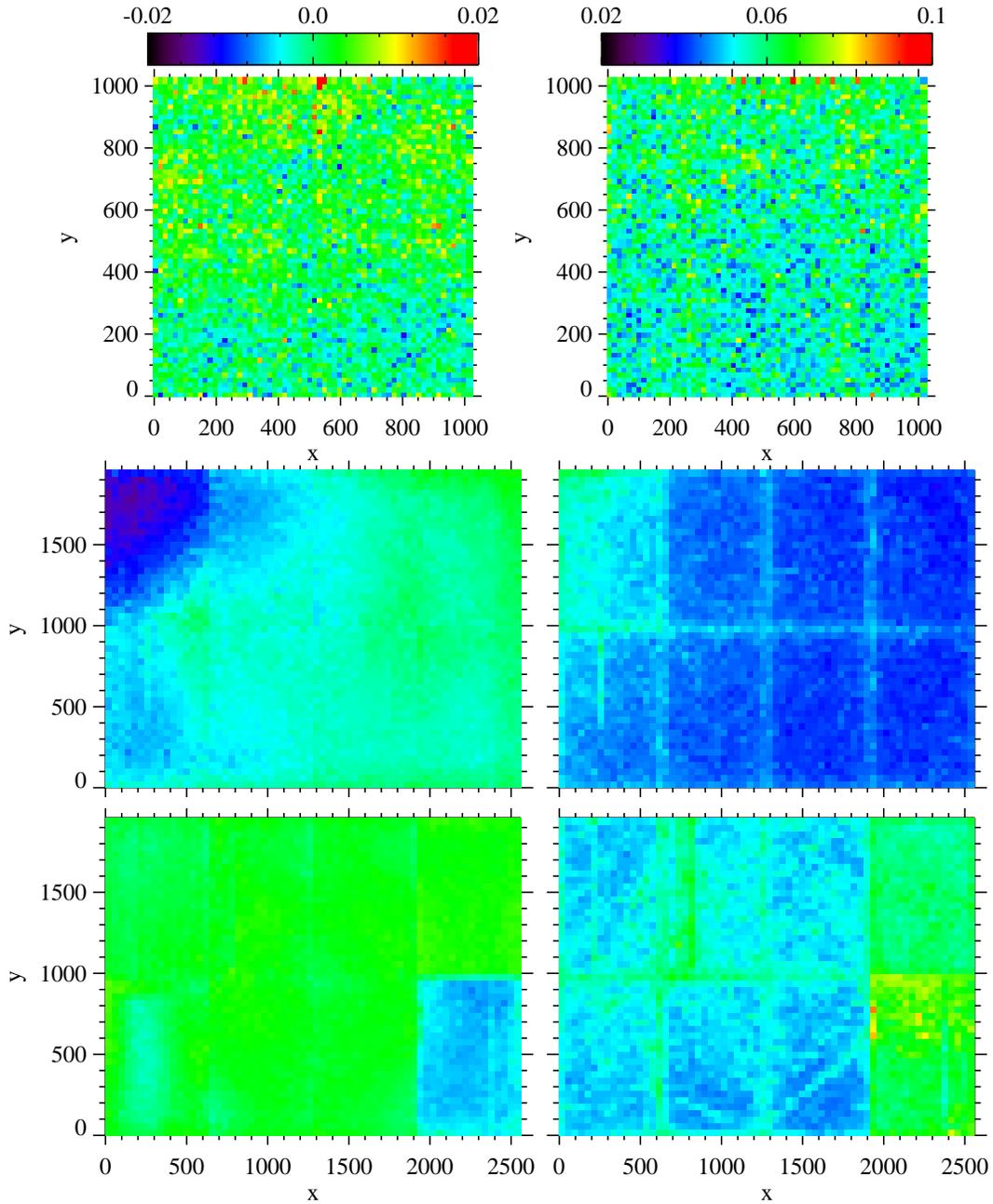}
\caption{
Similar to Fig.~\ref{xy_map_nosp}, but obtained after subtraction of ``super
flat-fields''. The median values are color-coded on a smaller range than in
Fig.~\ref{xy_map_nosp} (0.02 mag vs.~0.05 mag). On average, the medians are now
smaller than 0.005 mag, and the rms scatter is smaller than 0.05 mag. The higher
rms scatter indicates cells with sometimes problematic readouts.
\label{xy_map_sp}}
\end{figure}

We find super flat-field, $F(x,y)$, iteratively by first adopting $\alpha=1.0$
in Equation~\ref{mlinear} and $F(x,y)=0$ in Equation~\ref{zero_point}. For each
field, we determine $Z_0$, $Z_1$, $Z_2$, and $C$ by minimizing
Equation~\ref{chi_min} using calibration stars brighter than $m_{sdss} = 17$.
The super flat-field correction, $F(x,y)$, is then obtained by finding the
median of $m_{linear}-m_{sdss}$ residuals binned in $x-y$ pixels, and is shown
as a color-coded map in Figure~\ref{xy_map_nosp} ({\em left}) for each
observational setup. Figure~\ref{xy_map_sp} ({\em left}) shows the residuals
after the subtraction of Figure~\ref{xy_map_nosp} ({\em left}) maps. On average,
the median residuals are now smaller than 0.005 mag, compared to $\sim0.02-0.03$
mag before the correction was made.

\begin{figure}
\epsscale{0.7}
\plotone{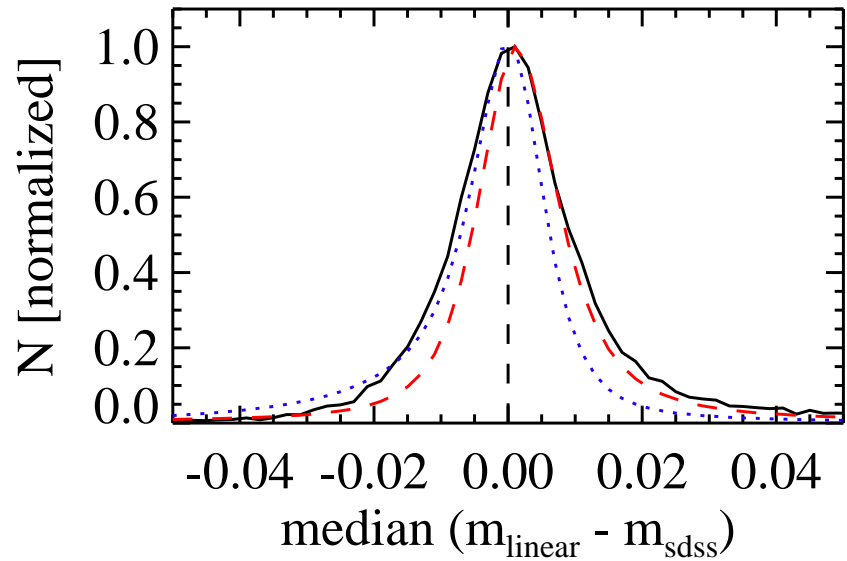}
\caption{
The medians of $m_{linear}-m_{sdss}$ residuals in $m_{sdss}$ bins for the small
({\em dots}), L1 ({\em crosses}), and L2 setup ({\em open triangles}). Apart
from slight non-linearity ($<0.02$ mag) in the L1 setup at bright magnitudes,
recalibrated LINEAR magnitudes are linear with magnitudes synthesized from SDSS
$gri$ photometry.
\label{dmag_msdss}}
\end{figure}

To verify our assumption of $\alpha=1.0$ (the linearity of instrumental
magnitudes), we adopt $\alpha=1.0$ in Equation~\ref{mlinear}, and determine
$Z_0$, $Z_1$, $Z_2$, and $C$ by minimizing Equation~\ref{chi_min} using
calibration stars brighter than $m_{sdss} = 17$. We bin $m_{linear}-m_{sdss}$
residuals in $m_{sdss}$ bins, and plot the medians as a function of $m_{sdss}$
in Figure~\ref{dmag_msdss}. A strong dependence of residuals on $m_{sdss}$ would
indicate non-linear ($\alpha \neq 1.0$) behavior of instrumental magnitudes.
With the exception of a small non-linearity ($<0.02$ mag) in the L1 setup for
$m_{sdss}<15$, the residuals do not depend on magnitude indicating that the
$\alpha=1.0$ assumption is valid.

To estimate the magnitude of correction introduced by including the
color-dependent term $C$ (accounts for the change in LINEAR bandpass due to
varying airmass) in Equation~\ref{mlinear}, we measure the rms scatter of
$C(g-i)$ values obtained from all calibration stars and fields. We find the rms
scatter to be $\sim0.02$ mag indicating that by correcting for this term the
systematic uncertainty in LINEAR photometry is reduced by $\sim0.02$ mag. On the
other hand, the inclusion of the $Z_0$ term is much more important as it reduces
the systematic uncertainty in LINEAR photometry by $\sim0.3$ mag (the rms
scatter of $Z_0$ values is $\sim0.3$ mag).

Once the color-dependent term, $C$, and the zero-point, $Z(x,y)$, are obtained
for a field, the LINEAR photometry for that field is recalibrated using
Equation~\ref{mlinear}. For ``orphan'' sources (LINEAR sources not associated
with SDSS objects) the color-dependent term is set to $C=0$, as $g-i$ colors
are not available for these sources. Therefore, the systematic uncertainty in
recalibrated photometry is slightly higher for orphans (by $\sim0.02$ mag).

\subsubsection{Checks on Recalibrated Photometry}

\begin{figure}
\epsscale{1.0}
\plotone{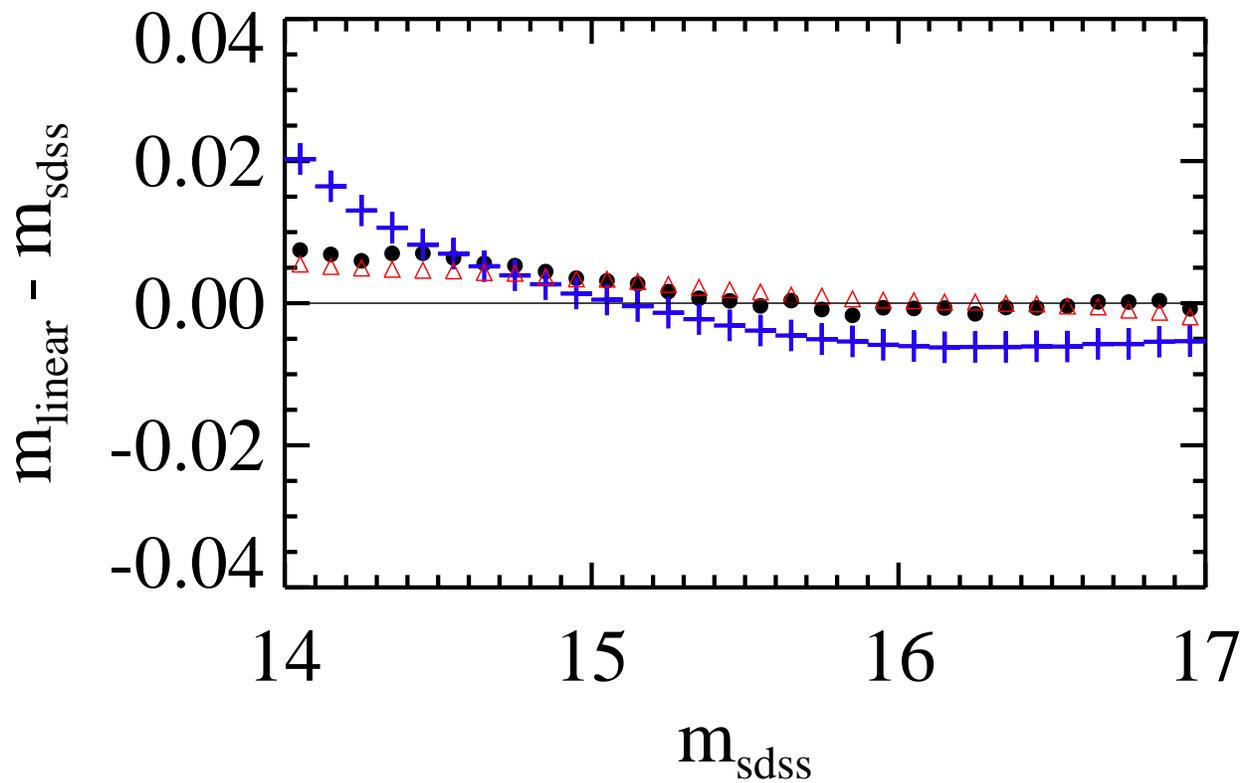}
\caption{
The distribution of zero-point offsets between SDSS and LINEAR systems for the
small ({\em solid line}), L1 ({\em dotted line}), and L2 setup ({\em dashed
line}), where the offset is measured for each field as the median of
$m_{linear}-m_{sdss}$ residuals. The distribution is about 0.01 mag wide,
implying that on average the LINEAR recalibration is about 0.01 mag offset from
the SDSS photometric system.
\label{median_dmag}}
\end{figure}

To quantify the quality of recalibration on a field-to-field basis, we calculate
$m_{linear}-m_{sdss}$ residuals for calibration stars, and find their median and
rms scatter (determined from the interquartile range) for each field. The median
of ($m_{linear}-m_{sdss}$) residuals per field is then an estimate of the
zero-point error. The distribution of these median offsets, shown in
Figure~\ref{median_dmag}, is about 0.01 mag wide and demonstrates the overall
quality of recalibrated photometry. About 10\% of fields are of low-quality (rms
scatter in $m_{linear}-m_{sdss}$ $>0.1$ mag), usually due to highly variable
cloud coverage. Sources detected in such fields are removed from further
consideration.

\begin{figure}
\epsscale{0.6}
\plotone{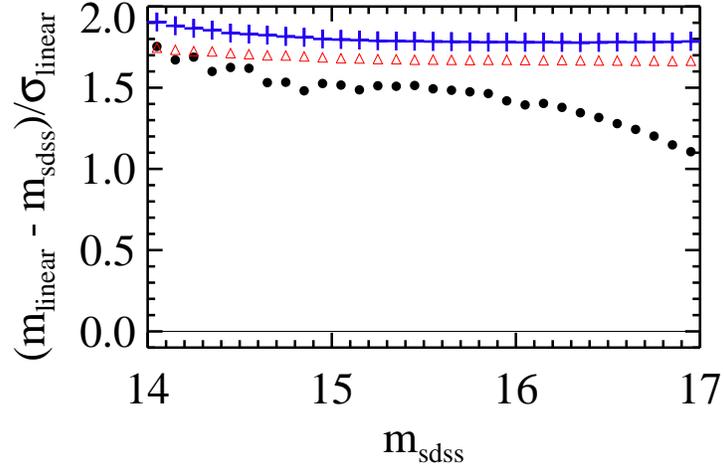}
\caption{
The rms scatter of $(m_{linear}-m_{sdss})/\sigma_{linear}$ values in $m_{sdss}$
bins for the small ({\em dots}), L1 ({\em crosses}), and L2 setup ({\em open
triangles}). The rms scatter values greater than 1 indicate underestimated
LINEAR magnitude errors. The pipeline-supplied errors for the small, L1, and
L2 setup are underestimated by about 40\%, 85\%, and 65\%, respectively.
\label{chi_msdss}}
\end{figure}

Another useful statistic is the rms scatter of $\chi$ values,
$(m_{linear}-m_{sdss})/\sigma_{linear}$, in $m_{sdss}$ bins. This rms scatter
should be equal to 1 and should not depend on $m_{sdss}$ if the
SExtractor-supplied magnitude errors ($\sigma_{linear}$) are correct Gaussian
errors. As shown in Figure~\ref{chi_msdss}, the rms scatter is greater than 1
indicating underestimated $\sigma_{linear}$ errors by up to a factor of $\sim2$.

We correct underestimated magnitude errors by multiplying them by 1.4, 1.85, and
1.65 for the small, L1, and L2 setup, respectively. In case of the small setup,
the factor of 1.4 is only the average value as SExtractor-supplied magnitude
errors for that setup seem to depend on magnitude. Therefore, the magnitude
errors for the small setup may be slightly overestimated or underestimated
depending on the brightness of the source. This is not a major issue as only a
small fraction of all LINEAR sources ($\sim0.1\%$) originate from the small
setup.

\begin{figure}
\epsscale{1.0}
\plotone{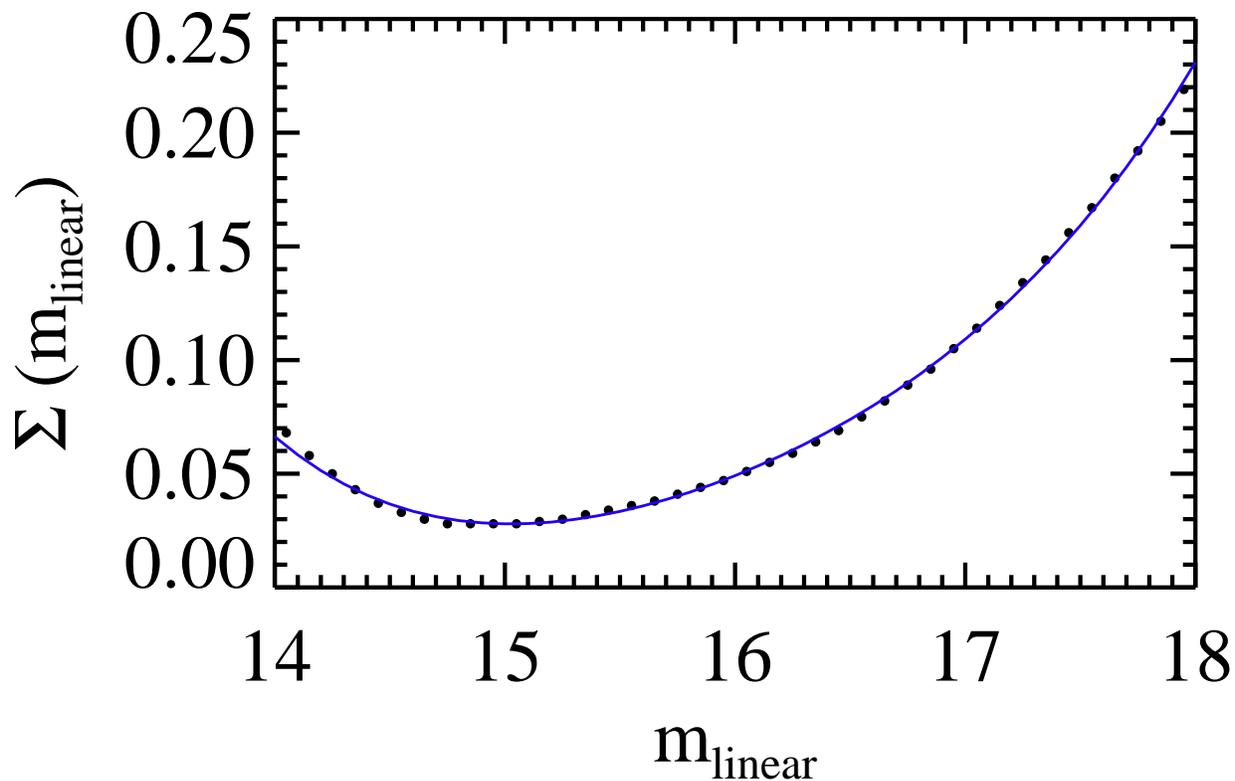}
\caption{
The median photometric error, $\Sigma$, as a function of magnitude. The median
photometric error is estimated as the median of $\sigma$ values in a magnitude
bin ({\em dots}), where $\sigma$ is the rms scatter of a time-series. The
analytic description of $\Sigma(m_{linear})$ is given by Eq.~\ref{photo_error}
({\em solid line}), and was obtained by fitting a fourth-degree polynomial to
medians. The median photometric error increases rapidly from about 0.03 mag
at 15 mag, where systematic errors dominate, to 0.2 mag at 18 mag, where the
errors are dominated by photon statistics.
\label{error_vs_mag}}
\end{figure}

The reason for this underestimation of SExtractor-supplied magnitude errors is
not clear. Non-inclusion of systematic errors (e.g., due to flat-fielding) in
$\sigma_{linear}$ cannot explain this behavior for L1 and L2 setups. If additive
systematic errors were the reason for this underestimation, then
$(m_{linear}-m_{sdss})/\sigma_{linear}$ should decrease with magnitude because
$\sigma_{linear}$ is expected to increase with magnitude (see
Figure~\ref{error_vs_mag}). Yet, no such decrease is seen in
Figure~\ref{chi_msdss} for L1 and L2 setups.

\clearpage

\section{Characterization of the Recalibrated LINEAR Dataset\label{characterization}}

\begin{figure}
\epsscale{1.0}
\plotone{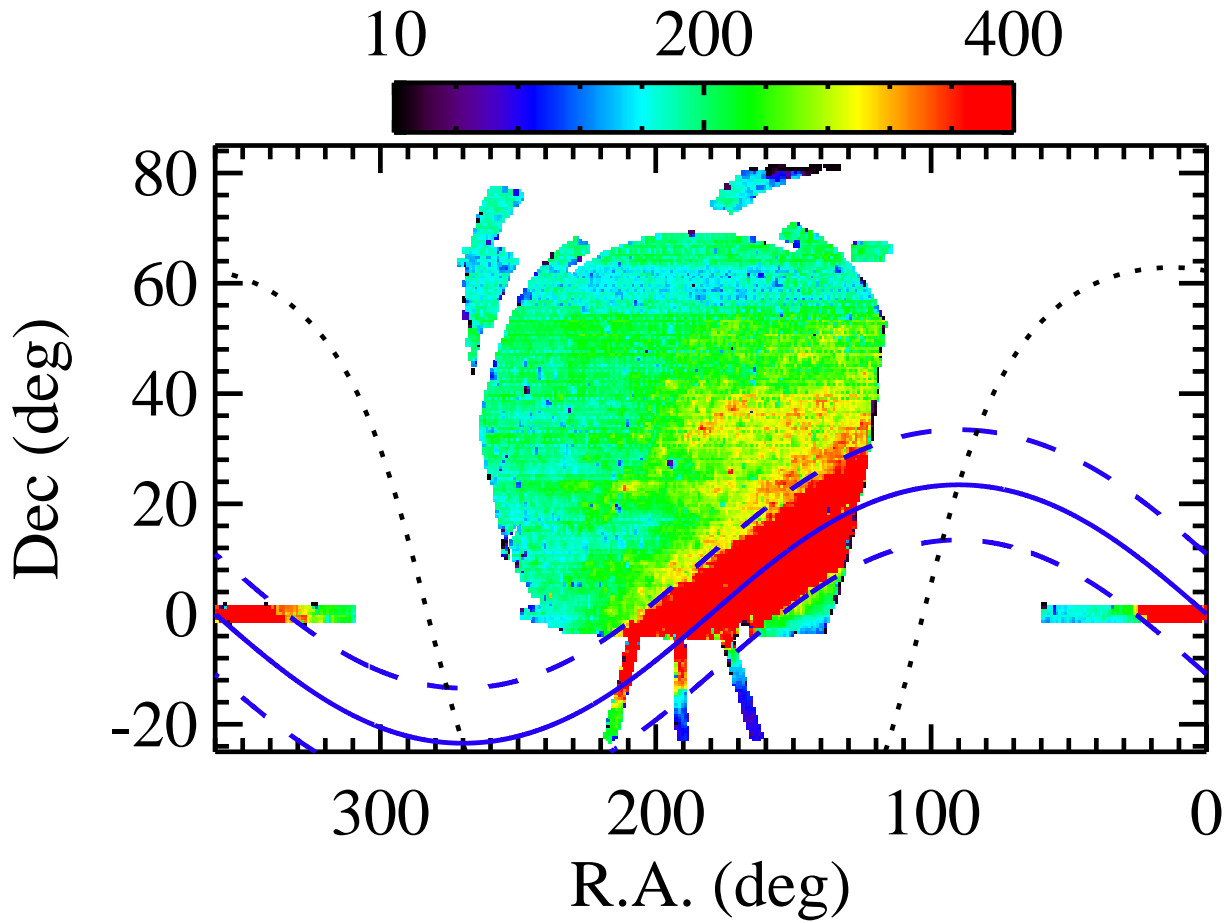}
\caption{
The median number of observations per object as a function of equatorial J2000.0
right ascension and declination coordinates. The values are color-coded
according to the legend, with values outside the range saturating. The dashed
lines show $\pm10\arcdeg$ of the ecliptic plane ({\em solid line}) and the
Galactic plane is shown as a dotted line. The average number of observations per
object within $\pm10\arcdeg$ of the ecliptic plane is $\sim460$, and $\sim200$
elsewhere.
\label{linear_coverage}}
\end{figure}

\begin{figure}
\epsscale{0.8}
\plotone{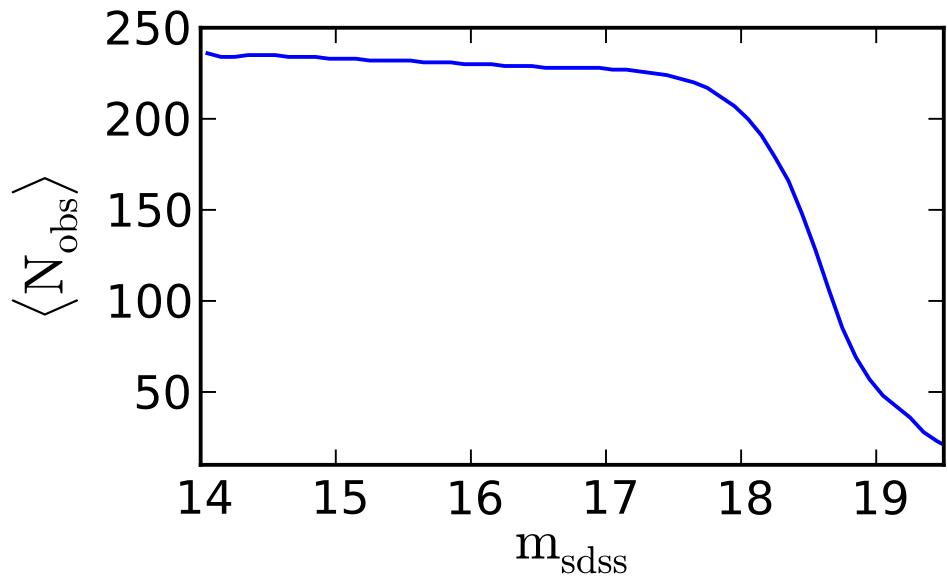}
\caption{
The median number of ``good'' observations (observations with flags$=$0) as a
function of $m_{sdss}$. The median number of ``good'' observations is $\sim200$
for $m_{sdss}<18$, after which it decreases precipitously (the $5\sigma$
detection limit is at $m_{sdss}\sim18$). About 10\% of detections have
flags$>0$.
\label{nobs_vs_mag}}
\end{figure}

There are about 5 billion sources in the recalibrated LINEAR dataset. These
sources were grouped into clusters using an implementation of the OPTICS
algorithm \citep{ank99} and a clustering radius of $2\arcsec$. There are about
25 million clusters with more than 15 observations. SDSS provides morphological
classification for 24 million clusters (7 million are resolved and 17 million
are unresolved by SDSS), and for 1 million of these clusters there is no SDSS
classification (they are slightly outside the SDSS footprint). Clusters with 15
or more sources are hereafter labeled as LINEAR objects or time-series. The
median number of epochs (observations) per object as a function position and
magnitude is shown in Figures~\ref{linear_coverage} and~\ref{nobs_vs_mag}.
The average number of epochs for objects brighter than 18 mag within
$\pm10\arcdeg$ of the ecliptic plane is $\sim460$, and $\sim200$ elsewhere.

Clusters with less than 15 sources are not tagged as objects. We have
statistically analyzed properties of 385 million sources ($\sim8\%$ of all
LINEAR sources) assigned to such clusters and have found that most of them are
very faint ($m_{linear}>19.5$), have very uncertain magnitudes
($\sigma_{linear}>0.4$ mag), and have high ellipticity (suggestive of cosmic
rays and image artifacts). Such clusters and their observations are not included
in our database. This also means that the LINEAR database does not include
observations of fast moving objects (i.e. objects moving faster than
$\sim1\arcsec$ day$^{-1}$).

\begin{figure}
\epsscale{1.0}
\plotone{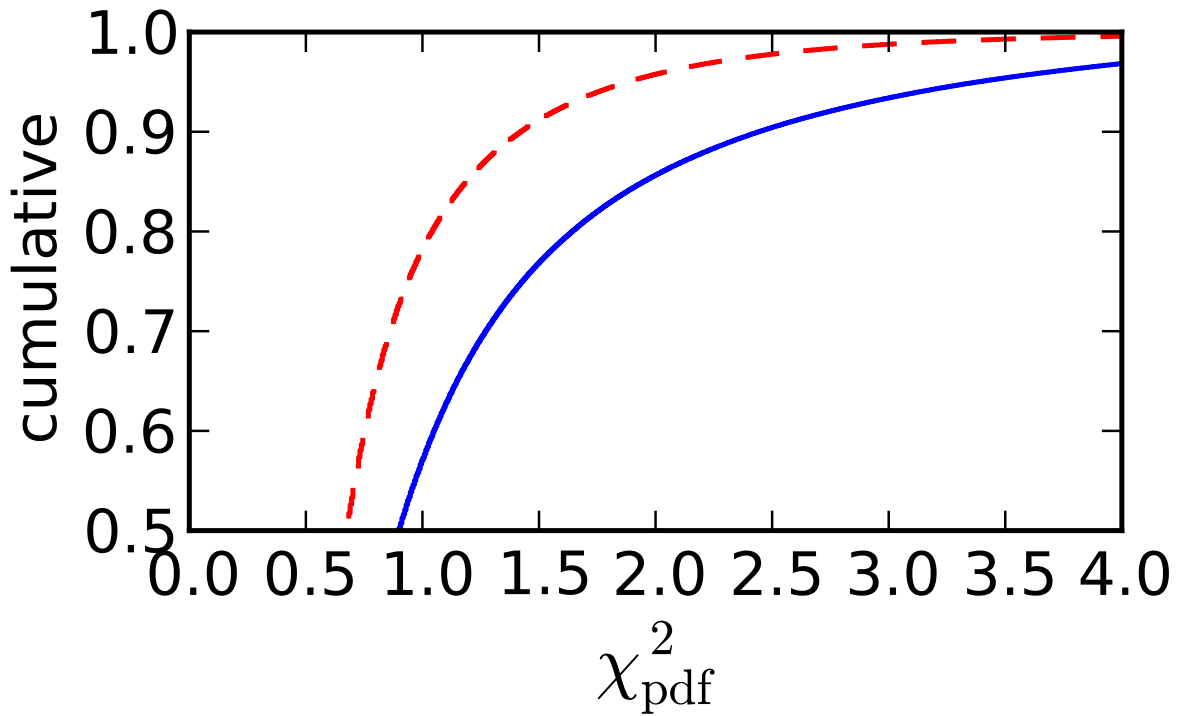}
\caption{
The cumulative distribution of $\chi^2_{pdf}$ values for a bright
($14.5<\langle m_{linear}\rangle<17$, {\em solid line}) and faint
($17<\langle m_{linear}\rangle<18$, {\em dashed line}) sample of LINEAR objects.
The median of distributions is $\sim1$, indicating that, on average, the
uncertainties in LINEAR magnitudes are well determined. The high-end tail of
distributions ($\chi^2_{pdf}>3$) is populated by truly and spuriously variable
objects (most likely dominated by the latter).
\label{chi2pdf_cum}}
\end{figure}

In order to enable discovery of variable objects and to further characterize the
LINEAR dataset, we have computed various low-order statistics for time-series,
such as the median recalibrated magnitude ($\langle m_{linear}\rangle$), rms
scatter ($\sigma$), $\chi^2$ per degree of freedom ($\chi^2_{pdf}$), skewness
($\gamma_1$), and kurtosis ($\gamma_2$). The cumulative distributions of
$\chi^2_{pdf}$ for LINEAR objects with more than 30 observations are shown in
Figure~\ref{chi2pdf_cum}.

The median of the $\chi^2_{pdf}$ distribution for bright LINEAR objects
($14.5<\langle m_{linear}\rangle<17$) is about 1, indicating that, on average,
the uncertainties in LINEAR magnitudes are well determined. For fainter LINEAR
objects ($17<\langle m_{linear}\rangle<18$), the median of the $\chi^2_{pdf}$
distribution is slightly lower ($\sim0.7$) indicating slightly overestimated
photometric errors.

The high-end tail ($\chi^2_{pdf}>3$) of the cumulative $\chi^2_{pdf}$
distribution suggests that $\sim5\%$ of bright LINEAR objects exhibit deviations
from the mean that are inconsistent with the Gaussian distribution of errors.
These objects could be variable or could have unreliable (non-Gaussian) errors.
While it is not impossible that some fraction of LINEAR objects are variable
($\sim1\%$ of sources brighter than $r=20$ are variable, see \citealt{ses06} and
references therein), a fraction of $\sim5\%$ seems unlikely. A more likely
explanation is spurious variability induced by non-Gaussian errors, caused by
blending of sources and unaccounted instrumental defects (e.g., bad pixels).

To estimate the dependence of the median photometric error ($\Sigma$) on
magnitude, we follow \citet{ses07} and calculate median $\sigma$ values in
$\langle m_{linear}\rangle$ bins. The dependence of median $\sigma$ on
$\langle m_{linear}\rangle$, shown in Figure~\ref{error_vs_mag}, can be modeled
as a fourth-degree polynomial
\begin{equation}
\Sigma(x) = 122.995-30.346x+2.8183x^2-0.116893x^3+0.0018295x^4,\label{photo_error}
\end{equation}
where $x=m_{linear}$. The smallest photometric errors are at $m_{linear}\sim15$
($\sim0.03$ mag) and are limited by systematic errors in calibration. The
increase in photometric error at brighter magnitudes is probably due to
saturation issues in LINEAR. The photometric error starts to increase rapidly
towards fainter magnitudes due to photon noise, and reaches $\sim0.2$ mag at
$m_{linear}\sim18$ (the adopted faint limit).

In addition to statistics calculated on light curves, we also calculated
time-averaged positions (median and mean right ascension and declination). To
estimate the reliability of time-averaged LINEAR positions, we compared them to
SDSS positions in $\sim1$ deg$^2$ pixels on the sky and calculated the median
difference in positions for each pixel. We find that, on average, the
time-averaged LINEAR positions are within $0.3\arcsec$ of SDSS positions, with
no systematic trends across the sky. The largest median difference (for a
$\sim1$ deg$^2$ pixel) is $\sim1\arcsec.5$. While this precision is sufficient
for most applications, we still recommend that SDSS positions are used whenever
possible to avoid patches where time-averaged LINEAR positions may be less
reliable.

\begin{figure}
\epsscale{0.8}
\plotone{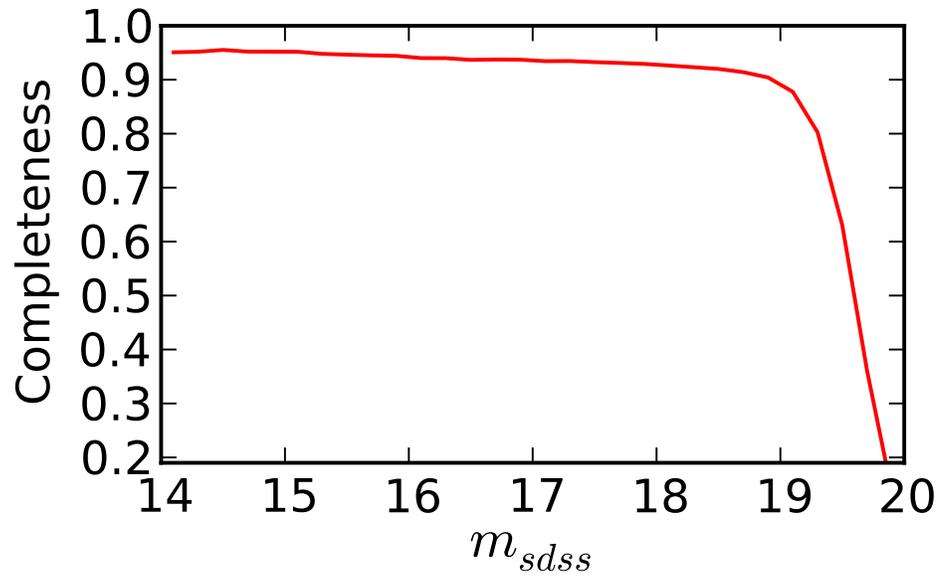}
\caption{
The completeness of the LINEAR catalog as a function of $m_{sdss}$ magnitude
(LINEAR magnitude synthesized from SDSS photometry, see Equation~\ref{msdss}).
The completeness is estimated as a fraction of SDSS objects matched to LINEAR
objects within $2\arcsec$. We find the object catalog to be more than 90\%
complete for $m_{sdss}<19$.
\label{completeness}}
\end{figure}

Since SDSS is about 4 mag deeper than LINEAR, it can be used to estimate the
completeness of the LINEAR object catalog (i.e., LINEAR object catalog is a
subset of the SDSS object catalog). To estimate the completeness of the LINEAR
object catalog, in Figure~\ref{completeness} we plot the fraction of SDSS
objects matched to LINEAR objects as a function of $m_{sdss}$ (LINEAR magnitude
synthesized from SDSS photometry, see Equation~\ref{msdss}). For this purpose we
use SDSS and LINEAR objects with $0<g-i<2.5$ (color range where
Equation~\ref{msdss} is valid), $170\arcdeg < {\rm R.A.} < 220\arcdeg$, and
$15\arcdeg < {\rm Dec} < 65\arcdeg$. As shown in Figure~\ref{completeness}, the
LINEAR {\em object} catalog is more than 90\% complete for $m_{sdss}<19$. This
estimate may be a bit misleading as the criterion for object creation is having
at least 15 detections. Even faint objects ($18<m_{sdss}<19.5$) are likely to be
detected at least 15 times out of an average of 200 observations (see
Figure~\ref{nobs_vs_mag}), if the observing conditions are better than average
(e.g, due to better seeing and darker skies).

\section{Preliminary Analysis of Variable Objects Selected from the LINEAR Dataset\label{analysis}}

\begin{figure}
\epsscale{1.0}
\plotone{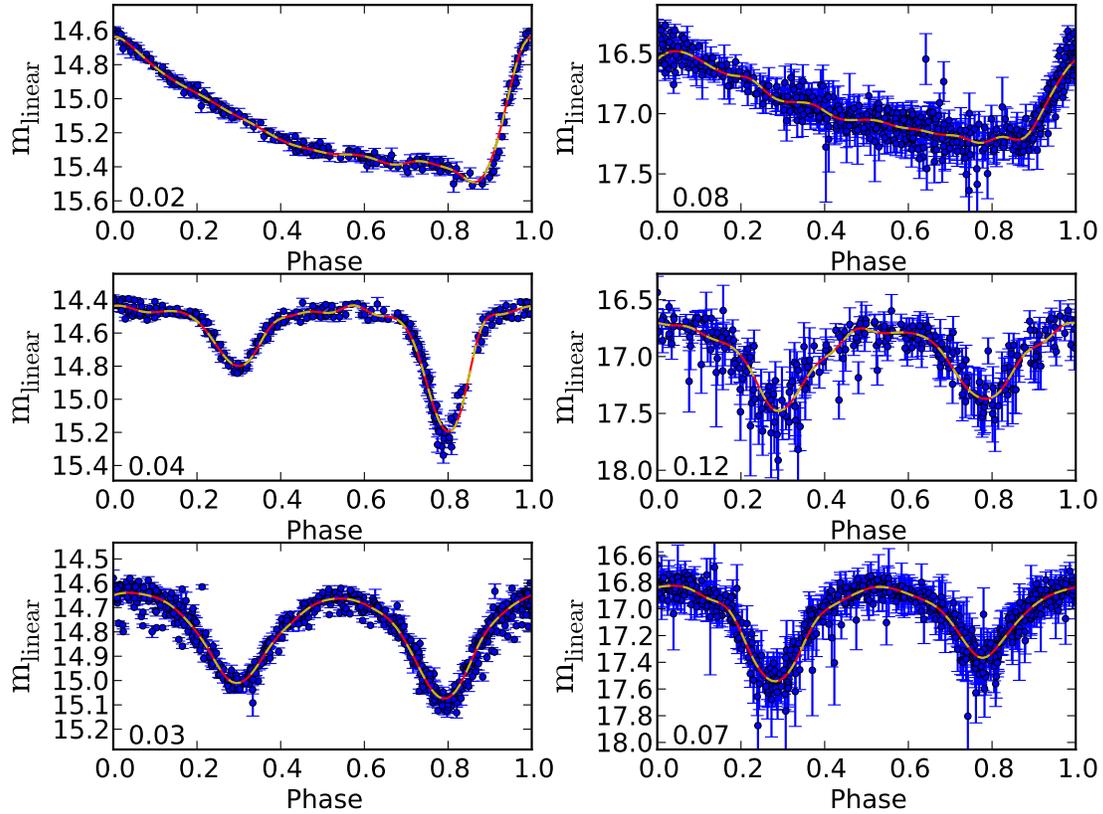}
\caption{Examples of period-folded light curves from the LINEAR dataset. The
left panels show bright ($\langle m_{linear}\rangle\sim15$) and right panels
show faint objects ($\langle m_{linear}\rangle\sim17$). The two top light curves
are RR Lyrae stars, and other are eclipsing binary systems. The number in the
lower left corner in each panel indicates the rms scatter around the cubic
spline fit to the binned mean values of the light curve using 20 points
({\em dashed line}). The light curve features are well defined, thanks to dense
LINEAR sampling, even for faint stars.
\label{example_LCs}}
\end{figure}

We have examined a small subset of objects with $\chi_{pdf}^2>3$ ($\sim$5,000)
and identified a subset of periodically variable stars. We confirmed expectation
from the preceding Section that only a few percent of the $\chi_{pdf}^2>3$
subsample are convincing cases of periodic variability. As an illustration of
the quality of LINEAR light curves at the bright
($\langle m_{linear}\rangle\sim15$) and faint end
($\langle m_{linear}\rangle\sim17$), we show their period-folded light curves in
Figure~\ref{example_LCs}. The main strength of the LINEAR survey is easily
discernible in these plots: even at the faint end, the light curve features are
well defined thanks to dense LINEAR sampling.

To exploit the full potential of the LINEAR photometric database, an automated
selection and classification of variable objects is desirable. This desideratum
is not easy to accomplish because the fraction of spurious objects with
$\chi_{pdf}^2>3$ is an order of magnitude larger than the fraction of truly
variable objects. In order to enable the deployment of automated selection and
classification methods, we have undertaken an extensive program of visual light
curve classification. Details of this program are described in a companion paper
(Palaversa, L. et al. 2011, in preparation, hereafter Paper II) and here we only
provide a brief description of the variable candidate selection and their
distribution in various color-color diagrams.

\begin{figure}
\epsscale{1.0}
\plotone{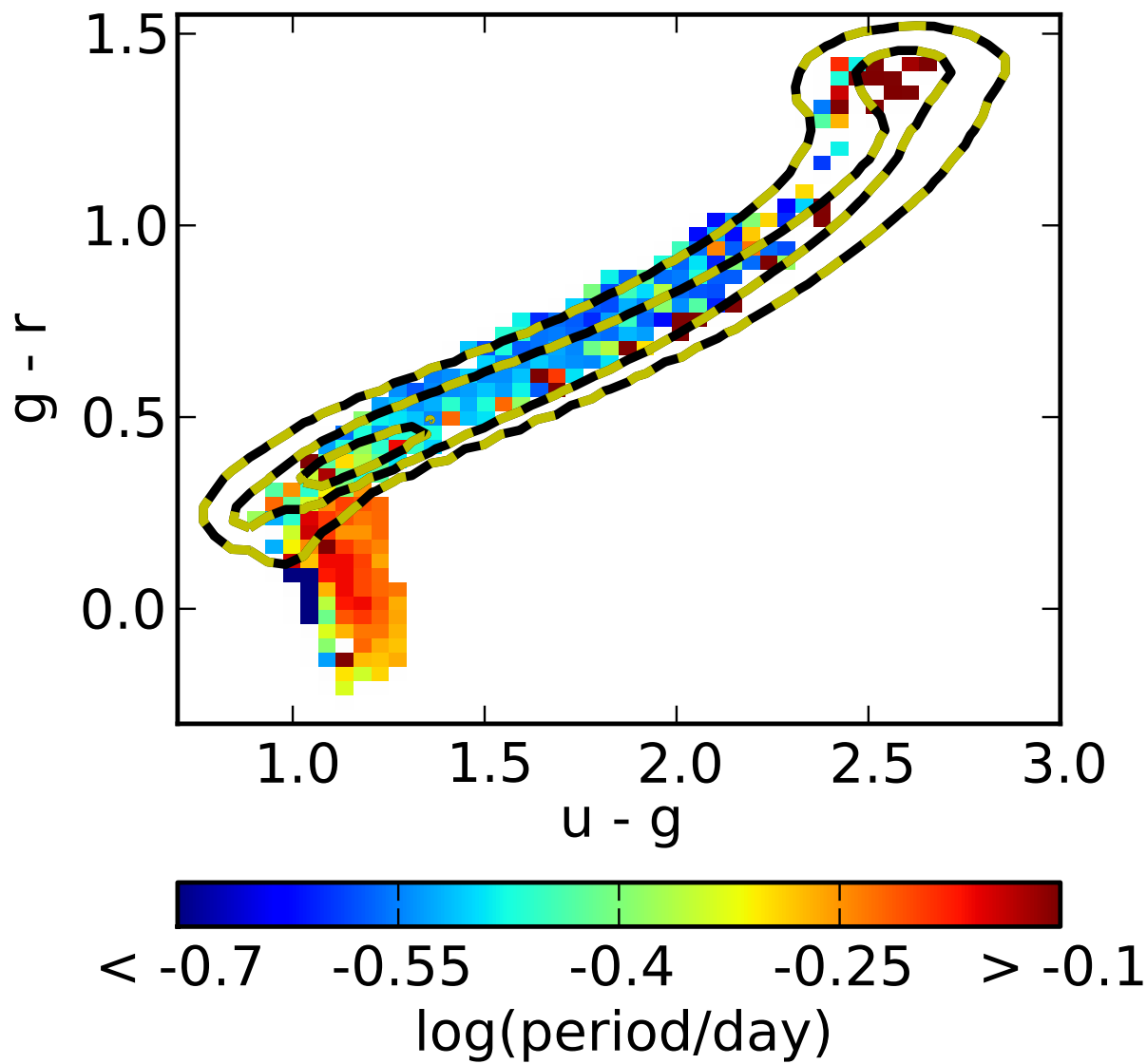}
\caption{
The distribution of $\log(period)$ for $\sim7000$ LINEAR variables in the SDSS
$u-g$ vs.~$g-r$ color-color diagram. The sample is limited to $14.5<r<17$ and
variability was confirmed by visual classification of light curves. Sources are
binned in 0.05 mag wide color bins, and the median values are color-coded
according to the legend on the right (values outside the range saturate in blue
or red). Contours outline the overall distribution of SDSS stars (with
$14.5<r<17$) at 1\%, 10\%, and 50\% of the peak density.
\label{ug_gr_logP}}
\end{figure}

In the first step, we define a flux-limited sample of objects with
$14.5<\langle m_{linear}\rangle<17$ and select about 200,000 light curves with
$\chi^2_{pdf}>3$ and $\sigma>0.1$ mag ($\sigma$ is the rms scatter in the light
curve). We limit classification to objects exhibiting periodic variability and
use phased light curves for visual inspection. The periods are determined using 
the Supersmoother algorithm \citep{rei94}. In total, about 7,000 LINEAR objects 
(3.5\% out of 200,000) were visually confirmed as periodic variables, with about
two thirds classified as RR Lyrae type $ab$ stars and eclipsing binary stars of 
all types. The remaining 1/3 include SX Phoenicis and Delta Scuti stars,
long-period variables, as well as possible RR Lyrae type $ab$ and eclipsing
binary stars with less certain classification. The distributions of classified
objects in the SDSS $u-g$ vs.~$g-r$ color-color diagram is shown in
Figure~\ref{ug_gr_logP}.

The differences between the distributions of all sources and those of the
confirmed variable subsample demonstrate that the latter are robustly classified
(i.e., the classification is not random). The most obvious difference is a much
higher fraction of RR Lyrae stars in the classified sample ($u-g\sim1.15$,
$g-r<0.3$). The eclipsing binary systems of all types dominate the visually
classified sample in the main stellar locus region. There seems to be a hint for
the median period of eclipsing binaries becoming longer for bluer stars. The
distribution of eclipsing stars is offset from the distribution of all stars in
the direction perpendicular to the stellar locus (offset to upper left). There
is also a lack of variable (eclipsing) systems of the late K and early M
spectral type. These results are discussed in more detail in Paper II.

\section{Summary\label{summary}}

In this work we have described astrometric and photometric recalibration of
data obtained by the asteroid survey LINEAR. The public access to the
recalibrated LINEAR dataset will be provided through the SkyDOT Web site 
(\url{http://skydot.lanl.gov}). A description of various cataloged parameters
that are available in this database is given in
Appendix~\ref{appendix:DBschema}.

The preliminary astrometry produced by the survey has been recalibrated using
the Astrometry.net software and USNO-B catalog as the reference astrometric
catalog. The precision (rms scatter) of single-epoch recalibrated LINEAR
coordinates is $\sim0\arcsec.6$ with respect to the reference (USNO-B)
astrometric catalog. The precision of time-averaged LINEAR coordinates is
$\sim0\arcsec.3$ with respect to SDSS positions.

In order to recalibrate the LINEAR photometry, we positionally matched the
LINEAR dataset to the SDSS DR7 imaging catalog and used SDSS as a catalog of
photometric standard stars. LINEAR instrumental magnitudes were corrected for
changes in the LINEAR bandpass (due to varying airmass) and for zero-point
variations across the CCD (due to imperfect flat-fielding). These corrected
LINEAR magnitudes were then offset on a per-LINEAR field basis to match the SDSS
magnitude zero-point system. In general, the recalibrated LINEAR magnitudes
($m_{linear}$) behave linearly, with the exception of a small non-linearity
($<0.02$ mag) in the L1 setup at magnitudes brighter than 15 mag.

We find the average uncertainty in recalibrated LINEAR magnitudes to be
$\sim0.03$ mag at the bright end ($m_{linear}\sim15$, the systematic
uncertainty) and $\sim0.2$ mag at the faint end ($m_{linear}\sim18$, the adopted
faint limit for the survey). The SExtractor-supplied magnitude errors were
found to be underestimated by $40\%-85\%$, independent of magnitude, and were
subsequently corrected for this behavior using multiplicative factors.

There are about 5 billion photometric measurements in the recalibrated LINEAR
dataset. The sources detected in each observation were grouped into clusters
(time-series or objects) using a clustering radius of $2\arcsec$. There are
about 25 million clusters with at least 15 observations (7 million resolved and
17 million unresolved by SDSS, the rest without morphological classification).
The median number of epochs (observations) per object brighter than 18 mag and
within $\pm10\arcdeg$ of the ecliptic plane is $\sim460$, and $\sim200$
elsewhere. The completeness of the LINEAR object catalog (relative to the SDSS
imaging catalog) was estimated to be better than 90\% for sources brighter than
19 mag. 

LINEAR dataset represents a major new resource for studying the variability of
faint optical sources. With an average of 200 observations per object, LINEAR
data provide time domain information for the brightest 4 magnitudes of the SDSS
survey. At the same time, LINEAR significantly extends the deepest similar
wide-area variability survey, the Northern Sky Variability Survey, by 3 mag.

In order to help automated discovery of variable objects, we have computed
various low-order statistics for time-series (median recalibrated magnitude, rms
scatter, $\chi^2$ per degree of freedom, skewness, and kurtosis). The
$\chi^2_{pdf}$ distribution for bright LINEAR objects (brighter than 17 mag) is 
centered at $\sim1$ (indicating that, on average, the uncertainties in LINEAR
magnitudes are well determined), and has a long high-end ($\chi^2_{pdf}>3$)
tail. About 5\% of LINEAR objects have $\chi^2_{pdf}>3$; in a sample with
Gaussian errors, less than 0.01\% of objects are expected to have
$\chi^2_{pdf}>3$. This high-end tail is populated by truly variable objects, but
is most likely dominated by spuriously variable objects with unreliable
(non-Gaussian) errors (caused by, for example, unaccounted instrumental defects
such as bad pixels or by blended sources).

Since the fraction of spurious variable objects is at least a few times greater
than the fraction of truly variable objects, we have undertaken an extensive
program of visual light curve classification in order to enable future
deployment of automated methods. Details of this program are described in a 
companion paper (Palaversa, L. et al. 2010, in preparation).

\acknowledgments

B.~Sesar thanks NSF grant AST-0908139 to J.~G.~Cohen and NSF grant AST-1009987
to S.~R.~Kulkarni for partial support. \v{Z}.I. acknowledges support by NSF
grants AST-0707901 and AST-1008784 to the University of Washington, by NSF grant
AST-0551161 to LSST for design and development activity, and by the Croatian
National Science Foundation grant O-1548-2009. Partial support for this work was
provided by NASA through a contract issued by the Jet Propulsion Laboratory,
California Institute of Technology under a contract with NASA. The LINEAR
program is sponsored by the National Aeronautics and Space Administration (NRA
No.~NNH09ZDA001N, 09-NEOO09-0010) and the United States Air Force under Air
Force Contract FA8721-05-C-0002. Opinions, interpretations, conclusions, and
recommendations are those of the authors and are not necessarily endorsed by the
United States Government.This research made use of tools provided by
Astrometry.net.

\appendix

\section{LINEAR Observing Strategy\label{appendix:cadence}}

To search for asteroids, the LINEAR program attempts to cover the entire portion
of sky available from the site each lunar month (lunation), with repeat searches
within 10 to 15 degrees of the ecliptic plane. The program schedules searches on
both telescopes an average of 25 nights per lunation, taking off 4-5 nights
around the full moon. After accounting for weather and equipment problems, about
20 observing nights are achieved per lunation with each telescope. To minimize
stray light, the survey fields are kept about $20\arcdeg$ away from the moon
during the week before and after the full moon.

To efficiently cover the available sky at minimal airmass, each telescope is
assigned a strip of sky that is aligned with either the ecliptic or the equator
and is typical 3-4 fields of view ($4-5\arcdeg$) wide in declination or ecliptic
latitude, and 150 fields of view ($240\arcdeg$) long in right ascension or
ecliptic longitude. The search box is divided into 5-7 regions each containing
about 100 fields. The regions are searched starting with the westernmost region
at the beginning of astronomical twilight in the evening, and ending with the
easternmost region at the end of astronomical twilight in the morning. All of
the fields in a single region are imaged in sequence, and then the sequence is
repeated in that region 4 more times. The 5 images of a given field are combined
into a ``frameset'' which forms the input to a moving target detection algorithm
that distinguishes asteroids from stars and noise based on their motion. This
process of cycling through the fields in a region 5 times produces a temporal
separation of 15 to 20 minutes between successive images of the same patch of
sky.

On subsequent nights, the entire search area for each telescope is generally
stepped north or south by the height of the region in order to cover the entire
sky, though gaps are occasionally unavoidable due to missed nights.  Over the
course of one lunation the entire available sky is generally searched once, and
areas within about $10\arcdeg$ of the ecliptic are searched two or three times.
Because the monthly search regions are quite wide in right ascension, there is
significant overlap in coverage area on adjacent lunar months. Therefore, a
given patch of sky can be observed up to 8 months in a row before it passes into
the daytime portion of the sky for 4 months. After a full year the search
covers the entire sky north of declination $-30\arcdeg$ except for a small
circle north of declination $82\arcdeg$ where the telescope has poor settling
ability. LINEAR's pattern of sky coverage produces repeat photometry data for
stars on several timescales ranging from the 15-20 minute interval between
images within a frameset, to a few days between repeat visits during one
lunation, to the month-long timescale between lunar months, to the yearly.

\section{LINEAR Database Schema\label{appendix:DBschema}}

\begin{figure}
\epsscale{1.0}
\plotone{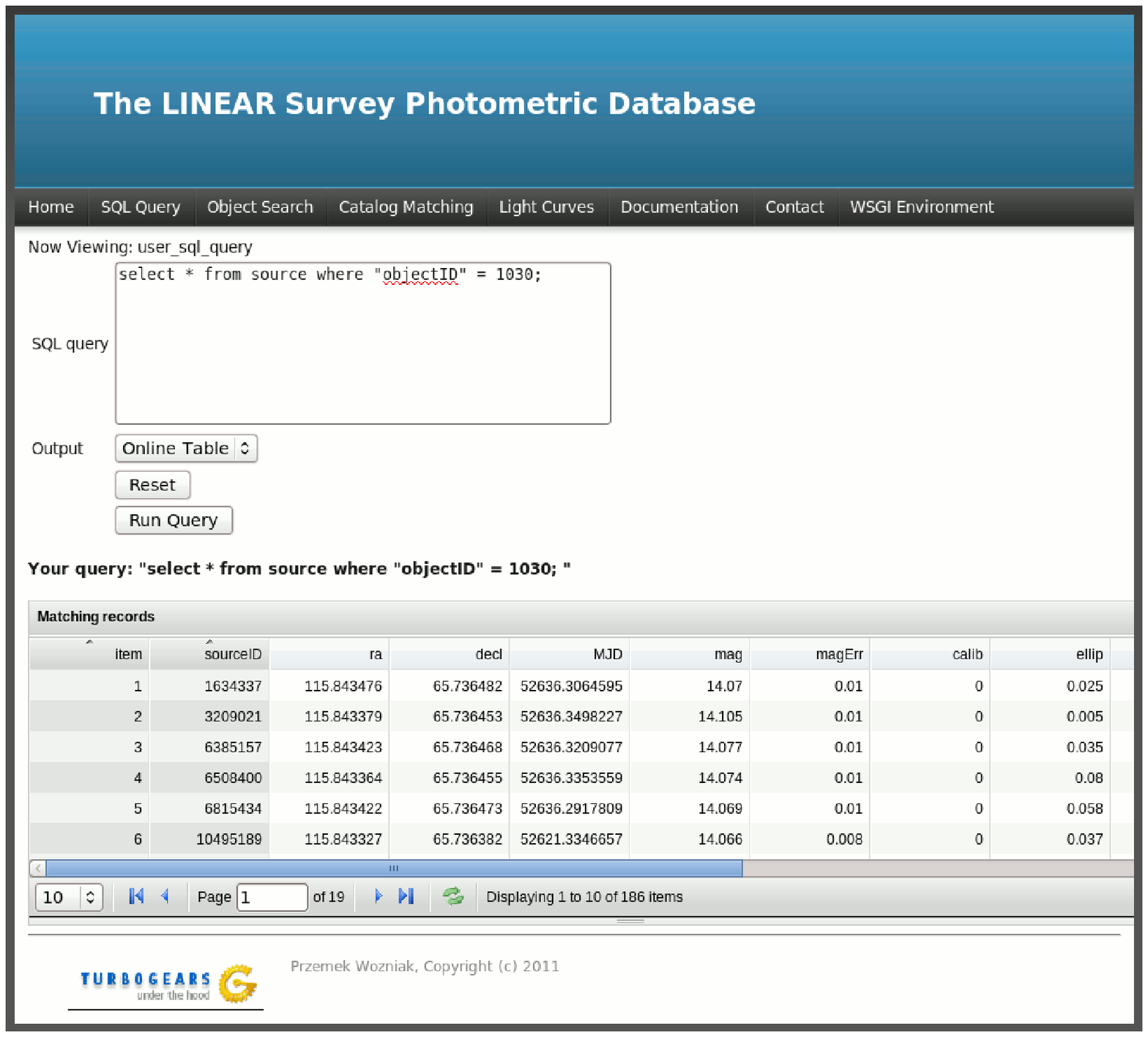}
\caption{
A screenshot showing the search interface for the LINEAR database.
\label{SkyDOT}}
\end{figure}

The LINEAR database hosted at SkyDOT contains two tables, ``Object'' and
``Source''. The ``Source'' table contains photometry and astrometry of
individual LINEAR detections (sources), while the ``Object'' table references
time-series obtained by clustering sources within $2\arcsec$. In addition, the
``Object'' table contains various low-level statistics calculated on time-series
and is cross-matched to external catalogs such as SDSS and 2MASS \citep{skr06}. A detailed description of various parameters provided by these tables is given
in Tables~\ref{source_schema} and~\ref{object_schema}, and a screenshot showing
the SkyDOT interface is shown in Figure~\ref{SkyDOT}.

\begin{deluxetable}{ll}
\tabletypesize{\scriptsize}
\tablecolumns{2}
\tablewidth{0pc}
\tablecaption{LINEAR ``Source'' Table Schema\label{source_schema}}
\tablehead{
\colhead{Field} & \colhead{Description} 
}
\startdata
sourceID & A unique number identifying this source \\
ra       & Equatorial J2000.0 right ascension \\
decl     & Equatorial J2000.0 declination \\
MJD      & Modified Julian Date \\
mag      & Recalibrated LINEAR magnitude \\
magErr   & Uncertainty in the recalibrated LINEAR magnitude \\
calib    & 1 if the source was used for calibration, 0 otherwise \\
ellip    & ``Ellipticity'' parameter measured by SExtractor \\
ximg     & Position of the source along CCD's $x$ axis \\
yimg     & Position of the source along CCD's $y$ axis \\
FWHM     & Full-Width-At-Half-Maximum measured by SExtractor \\
flags$^a$    & SExtractor internal flags \\
objectID & A number associating this source with an object in the ``Object'' table
\enddata
\tablenotetext{a}{See p.~28 in the SExtractor manual (\url{https://www.astromatic.net/pubsvn/software/sextractor/trunk/doc/sextractor.pdf}).}
\end{deluxetable}

\clearpage

\begin{deluxetable}{ll}
\tabletypesize{\tiny}
\tablecolumns{2}
\tablewidth{0pc}
\tablecaption{LINEAR ``Object'' Table Schema\label{object_schema}}
\tablehead{
\colhead{Field} & \colhead{Description} 
}
\startdata
objectID      & A unique number identifying this object \\
ra            & Mean equatorial J2000.0 right ascension \\
decl          & Mean equatorial J2000.0 declination \\
ra\_std       & Rms scatter in right ascension (in units of arcsec) \\
decl\_std     & Rms scatter in declination (in units of arcsec) \\
ra\_median    & Median equatorial J2000.0 right ascension \\
decl\_median  & Median equatorial J2000.0 declination \\
nptsTotal     & Number of detections \\
nptsGood      & Number of detections with flags$==$0 \\
mag\_mean     & Mean recalibrated LINEAR magnitude \\
mag\_wmean    & Weighted recalibrated LINEAR magnitude (by magnitude errors) \\
mag\_median   & Median recalibrated LINEAR magnitude \\
mag\_std      & Rms scatter of recalibrated LINEAR magnitudes \\
mag\_wstd     & Weighted rms scatter of recalibrated LINEAR magnitudes (by magnitude errors) \\
mag\_skew     & Skewness of recalibrated LINEAR magnitudes \\
mag\_kurtosis & Kurtosis of recalibrated LINEAR magnitudes \\
mag\_chi2dof  & $\chi^2$ per degree of freedom around the mean magnitude \\
mag\_rchi2dof$^a$ & Robust $\chi^2$ per degree of freedom around the mean magnitude \\
\hline
theta\_sdss$^b$ & Distance in arcsec from the SDSS source \\
sdssID        & SDSS Data Release 7 objID parameter \\
sdss\_ra      & SDSS J2000.0 right ascension \\
sdss\_dec     & SDSS J2000.0 declination \\
objtype       & 6 if unresolved in SDSS imaging (i.e., a point source), 3 if resolved \\
rExt          & Extinction in the SDSS $r$-band \\
uMod          & SDSS $u$-band model magnitude \\
gMod          & SDSS $g$-band model magnitude \\
rMod          & SDSS $r$-band model magnitude \\
iMod          & SDSS $i$-band model magnitude \\
zMod          & SDSS $z$-band model magnitude \\
uErr          & SDSS $u$-band model magnitude error \\
gErr          & SDSS $g$-band model magnitude error \\
rErr          & SDSS $r$-band model magnitude error \\
iErr          & SDSS $i$-band model magnitude error \\ 
zErr          & SDSS $z$-band model magnitude error \\
ug            & SDSS $u-g$ color (corrected for ISM extinction, $u-g=(uMod-1.87rExt) - (gMod-1.38rExt)$) \\
gr            & SDSS $g-r$ color (corrected for ISM extinction, $g-r=(gMod-1.38rExt) - (rMod-1.00rExt)$) \\
ri            & SDSS $r-i$ color (corrected for ISM extinction, $r-i=(rMod-1.00rExt) - (iMod-0.76rExt)$) \\
iz            & SDSS $i-z$ color (corrected for ISM extinction, $iz = (iMod-0.76rExt) - (zMod-0.54rExt)$) \\
gi            & SDSS $g-i$ color (corrected for ISM extinction, $g-i=(gMod-1.38rExt) - (iMod-0.76rExt)$) \\
pmL$^c$       & Proper motion in the Galactic latitude direction (in units of mas/yr) \\
pmB$^c$       & Proper motion in the Galactic longitude direction (in units of mas/yr) \\
pmErr$^c$     & Proper motion error (in units of mas/yr) \\
isolated      & 1 if SDSS CHILD flag is set to 0 (isolated$==$0 may indicate a close neighbor) \\
sdss\_satur   & 1 if SDSS SATURATED flag is set to 1 \\
sdss\_flags   & SDSS flags encoded as a 64-bit integer number \\
\hline
theta\_2mass$^d$  & Distance in arcsec from the 2MASS source \\
J             & 2MASS $J$-band magnitude \\
JErr          & 2MASS $J$-band magnitude error \\
H             & 2MASS $H$-band magnitude \\
HErr          & 2MASS $H$-band magnitude error \\
K             & 2MASS $K$-band magnitude \\
KErr          & 2MASS $K$-band magnitude error \\
jh            & 2MASS $J-H$ color (corrected for extinction, $j-h=(J-0.327rExt) -   (H-0.209rExt)$) \\
hk            & 2MASS $H-K$ color (corrected for extinction, $h-k=(H-0.209rExt) -   (K-0.133rExt)$) \\
jk            & 2MASS $J-K$ color (corrected for extinction, $(J-0.327rExt) -   (K-0.133rExt)$) \\
\enddata
\tablenotetext{a}{Calculated after eliminating 10\% of highest and 10\% of
lowest magnitudes.}
\tablenotetext{b}{This and other SDSS parameters set to NULL if no match within
$2\arcsec$ and not in SDSS footprint, set to 100 if in SDSS footprint but no
match within $2\arcsec$.}
\tablenotetext{c}{Obtained from the \citet{mun04} SDSS-USNOB proper motion catalog.}
\tablenotetext{d}{This and other 2MASS parameters set to NULL if no match within $2\arcsec$. Following \citet{cov07}, we only match to 2MASS sources with the following 2MASS flags: $rd_flag = 222$, $bl_flag=111$, and $cc_flag=0$. }
\end{deluxetable}

\clearpage


\clearpage

\clearpage

\clearpage

\clearpage

\clearpage

\clearpage

\clearpage

\clearpage

\clearpage

\clearpage

\clearpage

\clearpage

\clearpage

\clearpage

\clearpage

\end{document}